\documentclass[fleqn,usenatbib]{mnras}

\usepackage{newtxtext,newtxmath}

\usepackage[T1]{fontenc}

\DeclareRobustCommand{\VAN}[3]{#2}
\let\VANthebibliography\thebibliography
\def\thebibliography{\DeclareRobustCommand{\VAN}[3]{##3}\VANthebibliography}

\usepackage[dvipdfmx]{graphicx}	
\usepackage{amsmath}	

\newcommand{\spp}{{\rm p}}
\newcommand{\spe}{{\rm e}}
\newcommand{\sps}{{\rm s}}
\newcommand{\fric}{{\rm f}}
\renewcommand{\vec}[1]{\mbox{\boldmath $#1$}}
\newcommand{\exx}[2]{#1 \times 10^{#2}}       
\newcommand{\Dp}[2]{{\partial #1 \over \partial #2}}  
\newcommand{\DLp}[2]{{{\rm D} #1 \over {\rm D} #2}}  
\newcommand{\DDp}[2]{{\partial^2 #1 \over \partial {#2}^2}}  
\newcommand{\grad}{\nabla }                   
\renewcommand{\div}{\nabla \cdot }            
\newcommand{\B}{{\vec{B}}}                     
\newcommand{\E}{{\vec{E}}}                     
\newcommand{\A}{{\vec{A}}}                     
\newcommand{\J}{{\vec{j}}}                     
\newcommand{\V}{{\vec{v}}}                     
\newcommand{\W}{{\vec{w}}}                     
\newcommand{\uc}{{\vec{u}_{\rm c}}}            
\newcommand{\Surf}{{\cal{S}}}               
\newcommand{\taiseki}{{\cal{V}}}               

\usepackage{ulem}

\title[Angular Momentum from Pulsars]{On the Angular Momentum Extraction from the Rotation Powered Pulsars}

\author[Shibata \& Kisaka]{
Shinpei Shibata,$^{1}$\thanks{E-mail: shibata.shimpei@gmail.com (SS)}
Shota Kisaka$^{2}$
\\
$^{1}$Yamagata University Faculty of Science Graduate School of Science and Engineering, Kojirakawa 1-4-12, Yamagata, Yamagata 990-8560, Japan\\
$^{2}$Department of Physical Science, Hiroshima University, Higashi-Hiroshima 739-8526, Japan
}

\date{Accepted XXX. Received YYY; in original form ZZZ}

\pubyear{2021}

\begin{document}
\label{firstpage}
\pagerange{\pageref{firstpage}--\pageref{lastpage}}
\maketitle

\begin{abstract}
The rotation powered pulsar loses angular momentum at  a rate of the rotation power divided by the angular velocity $\Omega_*$.  This means that the length of the lever arm of the angular momentum extracted by the photons, relativistic particles and wind must be on average $c/\Omega_*$, which is known as the light cylinder radius.  Therefore, any deposition of the rotation power within the light cylinder causes insufficient loss of angular momentum.  In this paper, we  investigate  two  cases of this type of energy release: polar cap acceleration and  Ohmic heating in the magnetospheric current inside the star.
As  for the   first case, the outer magnetosphere beyond the light cylinder  is found to  compensate the insufficient loss of  the  angular momentum.  We argue that the energy flux coming from the sub-rotating magnetic field  lines must be larger than the solid-angle average value,  and as a result, an enhanced energy flux emanating beyond the light cylinder is observed in different phases in the light curve from those of emission inside the light cylinder. 
As for the second case, the stellar surface rotates more slowly than the stellar interior.  We find that the way the magnetospheric current closes inside the star is linked to how the angular momentum is transferred  inside the star.  We  obtain numerical solutions which shows that the magnetospheric current inside the star spreads over  the polar cap magnetic flux embedded in the star in such a way that electromotive force  is gained  efficiently. 
\end{abstract}




\section{Introduction}

Rotation powered pulsars (RPPs), by definition, release rotational energy 
at a rate of $\dot{E}=- \Im \Omega_* \dot{\Omega}_* $,
where $\Im$, $\Omega_*$, and $\dot{\Omega}_*$ are the moment of inertia, 
angular velocity and its time derivative of the neutron star.
The associated angular momentum $ \dot{L}= - \Im \dot{\Omega}_* $
is released  at the same time.
Therefore the ratio $ \dot{E}/\dot{L} =\Omega_* $ is unique.
This relation is usually not explicitly incorporated in theoretical modellings; 
it is justified on the basis of the fact that the relation is 
automatically satisfied as long as the system is solved self-consistently.
In reality, however, the magnetosphere of  RPPs is so complicated that
no models have ever treated the whole system  self-consistently.
Instead, some specific part of the magnetosphere or  neutron-star interior 
has been always  a subject of theoretical modelling, and accordingly 
the relation between the energy and angular momentum 
$ \dot{E}/\dot{L} =\Omega_* $  cannot be examined. 
Any of the past modelling works can be incompatible with 
the energy-angular-momentum relation.
In this paper, we  investigate two cases, 
the polar cap model and the current closure problem inside the neutron star with
Ohmic dissipation,
in which little angular momentum is lost.

As for the first case, the polar cap model,
rotational energy is extracted through photons
while hardly any angular momentum is  carried off by these photons.
Specifically when a curvature gamma-photon with energy $\epsilon$ is emitted
from the polar caps,
the  angular momentum that it carries off is at most
$\sim \epsilon  R_* / c$, 
whereas the value implied by the energy-angular-momentum relation
is  
$\epsilon  / \Omega_* = \epsilon  R_L / c$, 
where $R_*$ is the radius of the neutron star and
$R_L = c / \Omega_*$ is the light cylinder radius, i.e., 
the favoured length of the lever arm is $R_L$, which is much larger
than $R_*$.
In terms of classical electromagnetism, it is worth noting that
the wave zone of the magnetic dipole radiation originates around the light 
cylinder. In this case, the length of lever arm is $R_L$, regarding 
the wave front is not spherical but spirals in such a way that 
$\dot{E}/\dot{L}=\Omega_*$ holds.
Since the energy-angular-momentum relation is required
for the whole system,
even if the polar caps do not emit sufficient angular momentum ($R_* \ll R_L$),
a part of the outer magnetosphere 
must emit photons with a lever arm length larger than $R_L$
to compensate the insufficiency
or equivalently electromagnetic waves are modified 
beyond the light cylinder. 
This may lead to a hypothesis that 
the polar cap emission will not make any significant effect
on the magnetospheric structure.
However, the observed radio luminosity becomes
comparable with the spin-down luminosity for old pulsars
\citep{2020RAA....20..188W}.  
This  fact implies that the polar cap luminosity can 
contribute to a significant fraction of $\dot{E}$, and that 
the inefficiency of the angular momentum loss may cause some
crucial effect on the outer magnetosphere.

As for the second case, the Ohmic dissipation inside the star,
the magnetospheric current goes out of  and  comes back
to the star,  thereby conveying the rotation energy  of the star  
 away through the magnetosphere.
The magnetospheric current runs inside the star to make
the circuit closed.
This  current generates the braking torque.
Note that the term  ``magnetospheric current inside the star''  does not include
the current  generated through the magneto-thermal evolution of the neutron star
but means the current that is connected to the magnetosphere.
If the Ohmic dissipation takes place on the magnetospheric current inside the star,
a part of the rotational energy  is converted to heat, which
is  eventually radiated away, while there is no loss of angular momentum from the system.
This effect was not taken into account in the previous works of this subject 
\citep{{2015SSRv..191..207B}, {2019MNRAS.487.3333K}}
which considered how the magnetospheric current ran inside the  star.

In this paper, we investigate the above-mentioned two cases and
discuss what would happen when
the rotational energy is lost but not the angular momentum.
Especially, for the second case, we demonstrate
how the magnetospheric current inside the star is determined.

\section{Basic relations}

In this paper, 
we consider the obliquely rotating magnetosphere
in a steady state. 
In this case,
the electric field is written in the form of
\begin{equation} \label{torq.a-E}
\E = - {1 \over c} \uc \times \B - \grad \Phi
\end{equation}
without loss of generality \citep{1999ISMP...99.....M}, 
where $\uc = \vec{\Omega}_* \times \vec{r}$ is the corotation vector,
$\B$ is the magnetic field, 
and
$\Phi$ is a scalar potential called the ``non-corotational electric potential''.
The first term of the equation is the corotational electric field
$\E_{\rm c} = - \uc \times \B /c$.
The field-aligned electric field is given by 
$E_\parallel = - \hat{\B} \cdot \grad \Phi$,
where 
$\hat{\B}=\B / |\B |$ is the unit vector along the magnetic field.
Some useful formulae under the steadiness condition are given in the appendix.

Here let us assume at the moment that the star is a perfect conductor satisfying
the ideal-magneto-hydrodynamic (MHD) condition, $\E + \V \times \B /c =0$,
and rotates rigidly with $\V = \uc$,
where $\V$ indicates the velocity field. 
Then the electric field  is $\E_{\rm c}$ and $\Phi = \mbox{constant} = 0$
inside the star. If the surrounding space is filled with  plasma
satisfying the ideal-MHD condition, then 
\begin{equation}
{1 \over c }( \V - \uc )\times \B = \grad \Phi
\end{equation}
holds so that $\B \cdot \grad \Phi =0$, meaning that 
the value $\Phi =0$ on the stellar surface
propagates into the magnetosphere along $\B$.
Thus we have the corotation electric field and the iso-rotation
\begin{equation}
\V = \uc + \kappa \B
\end{equation} 
in the magnetosphere, where $\kappa$ is a scalar function.

The torque density in the star is  given by
$(\uc / \Omega_* ) \cdot (\J \times \B /c)$,
and therefore the net torque on the star is
\begin{equation} \label{torq.a-N}
N 
= {1 \over \Omega_* } \int_{\taiseki_*} \uc \cdot \left( {1 \over c} \J \times \B \right) \ dV
= {1 \over \Omega_* } \int_{\taiseki_*} \E_{\rm c} \cdot \J \ dV, 
\end{equation}
where $\J$ is the current density of the magnetospheric current, and
$\taiseki_*$ is the stellar volume.
The rotational power released from the star is thus given by
$\Omega_* N$. 
The Poynting theorem
\begin{equation}
\Dp{U}{t}+\div \vec{S} = - \J \cdot \E
\end{equation}
where
\begin{equation}
U= { E^2 + B^2 \over 4 \pi }, \ \ \ 
\vec{S} = {c \over 4 \pi } \E \times \B ,
\end{equation}
reduces to
\begin{equation} \label{torq.a-Poynting}
\div ( \vec{S} - U \uc ) = - \J \cdot \E 
\end{equation}
in the steady condition.
Combining (\ref{torq.a-N}) and (\ref{torq.a-Poynting}) with
the assumption $\E = \E_{\rm c}$,
we have the rotation power represented by the surface integral 
over the stellar surface ${\Surf}_*$:
\begin{equation}
\dot{E} = -\Omega_* N =
\int_{{\Surf}_*} ( \vec{S} - U \uc ) \cdot \ d \A .
\label{torq.a.Edot}
\end{equation}
This indicates that
the rotational energy flowing out to the magnetosphere is accounted
by the Poynting flux on the stellar surface.

\section{Effect of the Polar Cap Accelerator on the Global Structure}
\label{magsec}

We investigate the influence of the polar cap acceleration on the
global torque balance of the magnetosphere.

\begin{figure}
\begin{center}
\includegraphics[width=10cm,page=1]{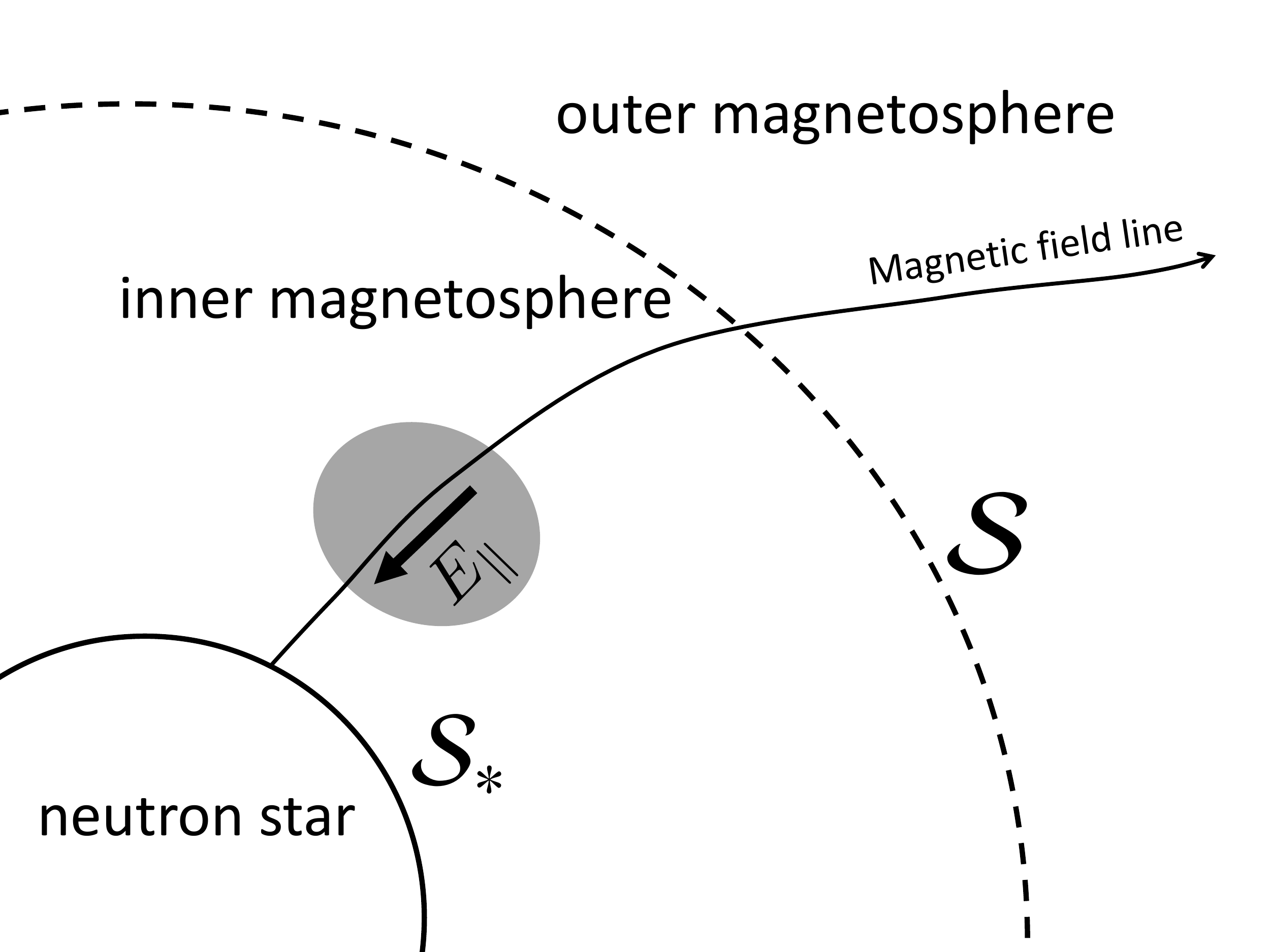}
\caption{ \label{torq-fig34} A simple model of a magnetosphere 
with a field-aligned particle accelerator. 
Circular solid-line $\Surf_*$ is the stellar surface and the
dashed-curve ${\Surf}$ is an arbitrary closed surface.
Gray-shaded region indicates a particle-acceleration region.}
\end{center}
\end{figure}

We use a simple model as described below 
and illustrated in Fig.~\ref{torq-fig34}.
Suppose a closed surface $\Surf$ surround the neutron star.
The ideal-MHD condition is satisfied inside the surface $\Surf$
except for 
a particle-acceleration region along open field lines 
(the grey region in the figure). 
The region between $\Surf$ and ${\Surf}_*$ may be called the ``inner magnetosphere'',
provided that $\Surf$ is within the light cylinder.
The region beyond $\Surf$ may be called the ``outer magnetosphere'', where
the ideal-MHD condition is assumed to hold.
The outer magnetosphere possesses a field-aligned outflow, the pulsar wind.

Immediately above the stellar surface ${\Surf}_*$ and below the acceleration region, 
$\Phi =0$, 
$\E = \E_{\rm c}$
and the flow obeys the relation $\V = \uc + \kappa \B$.
In the acceleration region, by contrast, 
there is a field-aligned electric field
$E_\parallel = - \hat{\B} \cdot \grad \Phi \not= 0$.
Even if the ideal-MHD condition is recovered beyond the acceleration
region,
variation of the non-corotational electric potential causes drift motions: 
\begin{equation} \label{torq.a-drift}
\V = \uc - { c \B \times \grad \Phi \over B^2 } + \kappa^\prime \B .
\end{equation}
This may be  attributed to the sub-pulse drift motion 
\citep{2020ApJ...896..168S}. 

The equation of motion for the particles in the inner magnetosphere
may be
\begin{equation} \label{eqmo1}
{1 \over c } e_\sps n_\sps \V_\sps \times \B + e_\sps n_\sps \E =
n_\sps \left( \Dp{}{t} + \V_\sps \cdot \grad \right) (m_\sps \gamma_\sps \V_\sps)  
- \vec{F}_{\rm ext}
\end{equation}
where $\vec{F}_{\rm ext}$  represents any non-electromagnetic forces,
the subscript ``s'' indicates the particle species, and 
$e_\sps$, $n_\sps$, $\V_\sps$, and $\gamma_\sps$ are respectively
the charge, number density, velocity, and the Lorentz factor.
For the reaction force by the curvature radiation we have
\begin{eqnarray}
\vec{F}_{\rm ext} &=& 
- {n_\sps {\cal P}_\sps  \over c} {\V_\sps \over c} , 
\\
{\cal P}_\sps &=& (2 e_\sps^2 / 3 c^3) \gamma_\sps^4 
                |\V_\sps \times (\nabla \times \V_\sps ) |^2
= (2 e_\sps^2 c / 3 R_{\rm c}^2 ) \gamma_\sps^4 ,
\end{eqnarray}
where $R_{\rm c}$ is the the radius of curvature.
The work done by the electromagnetic force per unit time is
given by the left-hand side of 
$\V_\sps \cdot$(\ref{eqmo1}) for each particle species, so that in total we have
\begin{equation} \label{torq-edotin}
\dot{E}_{\rm in} = 
\int_{\taiseki_{\rm in }}\sum_\sps \V_\sps \cdot 
( e_\sps n_\sps \E ) d V 
= \int_{\taiseki_{\rm in }} \J \cdot \E d V ,
\end{equation}
where $\taiseki_{\rm in}$ indicates the  volume between the surface
$\Surf$ and ${\Surf}_*$.
The value $\dot{E}_{\rm in}$ is the energy carried off by photons and/or
particles from the inner magnetosphere.

The component along the rotation axis of the
torque (the transferred angular momentum per unit time)  is given by
$(\uc / \Omega_*) \cdot$(\ref{eqmo1}):
\begin{eqnarray}
\dot{L}_{{\rm in}}
&=& {1 \over \Omega_* } \int_{\taiseki_{\rm in}} \sum_\sps \uc \cdot \left[
{1 \over c } e_\sps n_\sps \V_\sps \times \B + e_\sps n_\sps \E 
\right] d V \\
&=& { 1 \over \Omega_* }  \int_{\taiseki_{\rm in}} ( \J \cdot \E + \J \cdot \grad \Phi - q \uc \cdot \grad \Phi )\ dV ,
\label{torq-Ldotin}
\end{eqnarray} 
where  (\ref{torq.a-E})
is used in the last manipulation and 
$q=\sum_\sps e_\sps n_\sps$ is the charge density.
Subtracting 
$\Omega_*$ times (\ref{torq-Ldotin})
from 
(\ref{torq-edotin}) yields 
\begin{eqnarray} 
\label{torq-balancein}
\Delta \dot{E}_{\rm in} \equiv
\dot{E}_{\rm in}
- \Omega_* \dot{L}_{\rm in}
&=& - \int_{\taiseki_{\rm in}} (\J - q \uc ) \cdot \grad \Phi \ dV
\\
& \approx & \int_{\taiseki_{\rm in}} j_\parallel E_\parallel \ dV
 \not= 0 ,
\end{eqnarray}
where $j_\parallel = \J \cdot \hat{\B}$,
the approximation in the last step is based on
the facts that the ratio of the microscopic to macroscopic time scales
$|j_\perp / j_\parallel |= 
|\J \times \hat{\B}| /| \J \cdot \hat{\B}| 
\approx (c/R_{\rm c})/(e_\sps B/\gamma_\sps m_\sps c)$ 
is negligible,
and $|\uc | \ll c  $  holds in the inner magnetosphere. 
The derived equation indicates that no significant amount of
the angular momentum is lost from the inner magnetosphere.

In our simple model, 
the outer magnetosphere is treated in the one-fluid approximation 
with the ideal-MHD condition.
The equation of motion is
\begin{eqnarray} \label{torq-eqmowind}
{ 1 \over c} \J \times \B + q \E &=& 
n \left( \Dp{}{t} + \V \cdot \grad \right) (m \gamma \V) \\
&=&
n \left[ \Dp{m \gamma \V}{t} + \grad (mc^2 \gamma )
         - \V \times (\grad \times m \gamma \V) 
  \right] .
\end{eqnarray}
From 
$\V \cdot$ (\ref{torq-eqmowind})
and
$(\uc/\Omega_* ) \cdot$ (\ref{torq-eqmowind}),
it is straightforward to obtain the energy and angular momentum conservation laws:
\begin{equation}
\div \vec{\cal E} =0, \ \ \ \
\div \vec{\cal L} =0 ,
\end{equation}
where
\begin{eqnarray}
\vec{\cal E} &=&  
mn\gamma c^2 \W + \vec{S} - U \uc ,
\\
\vec{\cal L} &=& 
nm \gamma [(\uc / \Omega_* ) \cdot \V ] \W + (S - U \uc)/\Omega_* - (\J - q\uc )(\Phi/\Omega_* ) ,
\end{eqnarray}
where $\W = \V - \uc$.
The first terms of the  expressions above
are  the contribution from the particles,
while the last two terms are from the electromagnetic fields.
The energy-angular momentum outflow through the surface $\Surf$
can be evaluated  with the last two terms, namely 
\begin{eqnarray}
\label{torq.a-Eout}
\dot{E}_{\rm out}
&=& \int_{\Surf} (\vec{S} - U \uc ) \cdot d \A , \\
\label{torq.a-Lout}
\dot{L}_{\rm out}
&=& 
{1 \over \Omega_* } 
\int_{{\Surf}} (\vec{S} - U \uc ) \cdot d \A \nonumber \\
&& + {1 \over \Omega_* } \int_{{\Surf}} [(\J - q \uc )(- \Phi ) ] \cdot d \A .
\end{eqnarray}
These are used in part for the wind acceleration,
and the unused parts of them propagate
to the termination shock in  a  form of Poynting energy.
As is expected, 
(\ref{torq.a-Eout}), (\ref{torq.a-Lout}) and (\ref{torq-balancein}) are combined to yield
\begin{equation}
\dot{E}_{\rm in} + \dot{E}_{\rm out}
= \Omega_* ( \dot{L}_{\rm in} + \dot{L}_{\rm out}).
\end{equation}
 The torque balance 
 is satisfied in the whole system.
Although we are not concerned with what actually happens in the outer magnetosphere,
it is  certain that the outer magnetosphere carries angular momentum
so efficiently as to achieve the torque balance of the whole system.
This fact together with the source of Poynting flux being provided
by braking torque, (\ref{torq.a.Edot}), is included in classical
electromagnetism \citep{2010mfca.book.....B}.
If we define the effective angular velocity $\Omega_{\rm out}$
of the outer magnetosphere 
as $\Omega_{out} = \dot{E}_{\rm out} / \dot{L}_{\rm out}$,
we obtain the relation
\begin{equation}
\Omega_{\rm out} = 
 \Omega_* { 1 \over 
1 + 
\Delta \dot{E}_{\rm in}/ \dot{E}_{\rm out} 
}
< \Omega_* .
\end{equation}
This indicates that 
the effective light-cylinder radius $c/\Omega_{\rm out}$ is larger than $R_L$.
In other words,
the  lever arm of the angular momentum carried away from the outer magnetosphere
becomes larger
to compensate the inefficient angular momentum loss from the inner magnetosphere;
the body  inside the surface $\Surf$ is  effectively a slow rotator.
In this case,
if photons are emitted in the outer magnetosphere, the emission site
must be beyond the light cylinder.
Alternatively, the Poynting vector in the wind would
tilt  to satisfy $|\vec{r} \times \vec{S} |/ |\vec{S} | > R_L$.
What  exactly happens in the outer magnetosphere is discussed in \S{\ref{discuss}}.

\section{Effect of Ohmic Heating on the Torque Balance of the Star Interior}

If  Ohmic dissipation is present on the magnetospheric current which 
makes a circuit inside the star,
then  rotational energy is consumed without any loss of
angular momentum.
Let us consider this case  in this section.

For the present discussion
we employ a two-component (proton-electron) plasma with frictional force\footnote{%
The proton-electron model should be sufficient for the present discussion 
although inclusion of heavier-ion components would make the model closer to
the reality.}.
The equations of motion  can be written as 
\begin{eqnarray} 
        \label{torq-eqmo1}
m_\spp n_\spp \DLp{\V_\spp}{t} 
&=& e n_\spp \E + {1 \over c} e n_\spp \V_\spp \times \B + \vec{F}_\fric , \\
        \label{torq-eqmo2}
m_\spe n_\spe \DLp{\V_\spe}{t} 
&=& - e n_\spe \E -  {1 \over c} e n_\spe \V_\spe \times \B - \vec{F}_\fric , 
\end{eqnarray}
where the left-hand sides represent the Lagrangian time derivative, 
the subscripts  indicate the particle species, and
$\vec{F}_\fric$ is the frictional force which is given by
\begin{equation}
\vec{F}_\fric = - m_\spe n_\spe (\V_\spp - \V_\spe )\nu_{\rm coll}, 
\end{equation}
where $\nu_{\rm coll}$ is the effective collision frequency.
A set of equations
(\ref{torq-eqmo1}) and (\ref{torq-eqmo2}) is reduced to a familiar set of 
the one-fluid equation of motion and  Ohm's law:
\begin{eqnarray} 
\label{torq-eqmo3}
m_\spp n_\spp \DLp{\V}{t} 
& = & q \E + {1 \over c} \J \times \B , \\ 
\label{torq-eqmo4}
 \E + {1 \over c} \V \times \B & = & {\J \over \sigma} ,
\end{eqnarray}
where the electric conductivity 
$\sigma = e^2 n_\spe /m_\spe \nu_{\rm coll} $ is introduced and
the approximation is based on
$\V = (m_\spp n_\spp \V_\spp + m_\spe n_\spe \V_\spe )/(m_\spp n_\spp + m_\spe n_\spe ) \approx \V_\spp $,
$q =e(n_\spp-n_\spe)$, $\J=e(n_\spp \V_\spp - n_\spe \V_\spe)$ and
$q /en_e \ll 1$, whereas we ignore the Hall term, which is very small
for the magnetospheric current 
(see \S~\ref{discussB} for discussion).
Using  Ohm's law is equivalent to using the two-component model
with friction.

For the energy conservation,
adding 
(\ref{torq-eqmo1})$\cdot \V_\spp$
and
(\ref{torq-eqmo2})$\cdot \V_\spe$,
we have
\begin{eqnarray} \label{torq-enc}
m_\spp n_\spp \DLp{\V_\spp}{t} 
\cdot \V_\spp + 
m_\spe n_\spe \DLp{\V_\spe}{t} 
\cdot \V_\spe &=&  
\J \cdot \E + (\V_{\rm p} - \V_{\rm e} ) \cdot \vec{F}_\fric \nonumber \\
&=& \J \cdot \E - j^2 / \sigma ,
\end{eqnarray}
where the quasi-neutrality condition $n_e = n_p$ has been
applied, and $\J = e n_{\rm e} (\V_{\rm p} - \V_{\rm e})$
in the last manipulation.
Here we are concerned with only the rotational motion 
and  set
$\V_\spp = \Omega r \vec{e}_\varphi$, where
$r$ is the distance from the rotation axis and 
$\Omega$ is the angular velocity as a function of the position,  whereas
$\Omega_*$ is used for the angular velocity on the stellar surface.
We should note that $\Omega_*$ 
determines the ratio of the energy and the angular momentum
which flows out from the star. 
Ignoring the electron inertia,
we  obtain
\begin{equation} \label{torq-enc}
\Omega \dot{\Omega} m_\spp n_\spp r^2   
= \J \cdot \E - j^2 / \sigma .
\end{equation}
This is integrated over the volume of the star to give
\begin{equation} \label{torq-enc}
\langle \Omega \dot{\Omega} \rangle \Im   
= 
\int_{\taiseki_*} \J \cdot \E \ dV
 - \int_{\taiseki_*} (j^2 / \sigma) \ dV ,
\end{equation}
where
\begin{equation}
\Im = \int_{\taiseki_*} m_\spp n_\spp r^2 \ dV
\end{equation}
is the moment of inertia of the star, and the operator
$\langle \ \rangle$ indicates taking the average.
Equation  (\ref{torq-enc})  implies
that some part of the rotational energy goes into Joule heating
and the rest is the source of the Poynting flux which flows
into the magnetosphere; the volume integral of $\J \cdot \E$ 
is the surface integral of $\vec{S}$ on the stellar surface
(equations (\ref{torq.a-N}) and (\ref{torq.a.Edot})).

As for the component along the rotation axis of the angular momentum,
adding 
(\ref{torq-eqmo1})$\cdot ( \uc / \Omega_*)$
and 
(\ref{torq-eqmo2})$\cdot ( \uc / \Omega_*)$,
we have
\begin{equation}
\label{torq.a-Lstar}
r \vec{e}_\varphi \cdot \left(
m_\spp n_\spp \DLp{\V_\spp}{t} + m_\spe n_\spe \DLp{\V_\spe}{t} 
\right) 
= 
{1 \over \Omega_*} \left[
q \uc \cdot \E + {1 \over c} ( \J \times  \B ) \cdot \uc  
\right] , 
\end{equation}
where the frictional terms cancel each other out, 
following
the action-reaction principle, and the Ohmic term has no effect
on the angular momentum.
Equation (\ref{torq.a-Lstar}) is rewritten as, with
 $\E = - \uc \times \B /c - \grad \Phi$ 
and $\div (\J - q \uc )=0$,
\begin{eqnarray}
r \vec{e}_\varphi \cdot \left(
m_\spp n_\spp \DLp{\V_\spp}{t} + m_\spe n_\spe \DLp{\V_\spe}{t} 
\right) 
&=&
{1 \over \Omega_*} \left[
\J \cdot \E + 
(\grad \Phi ) \cdot (\J -q \uc )
\right] \nonumber \\ 
&=&
{1 \over \Omega_*} \left[ \J \cdot \E + \div \{(\J - q \uc ) \Phi  \} \right] .
\end{eqnarray}
The second term of the right-hand side is 
integrated over the star to  give zero
because $\Phi=0$ on the surface of the star:
\begin{equation}
\int_{\taiseki_*} \div \{(\J -q  \uc ) \Phi  \} dV =
\int_{{\Surf}_*} (\J -q \uc ) \Phi  \cdot  d \A = 0 .
\end{equation}
Consequently, we  obtain
\begin{equation}
\langle \dot{\Omega} \rangle \Im   
= 
{ 1 \over \Omega_* } 
\int_{\taiseki_*} \J \cdot \E \ dV .
\end{equation}
This means that all  the lost angular momentum goes into
the magnetosphere.

The ratio of the net losses of the energy and angular momentum 
is calculated to be
\begin{equation}
\langle \Omega \rangle \equiv
{ - \Im \langle \Omega \dot{\Omega} \rangle 
\over 
  - \Im \langle \dot{\Omega} \rangle }
= \Omega_* \left(
1 + {\int_{\taiseki_*} (j^2 / \sigma) dV
\over
\int_{\Surf_*} (\vec{S} - U \uc) \cdot d \vec{A} }
\right)
> \Omega_* .
\end{equation}
Therefore, the stellar interior rotates faster than the surface
when Ohmic dissipation takes place.
The main conclusion of this section is that
in determining
the magnetospheric current inside the star one must treat
the dynamics of the  interior matter, 
particularly the angular momentum transfer.

\section{ Distribution of the electric current Inside the Star} \label{currentdistribution}

In the previous section,
we  have found that the magnetospheric current inside the star is determined
not by  Ohm's law but by  magneto-hydrodynamics.
An illustrative example is given in this section.

Since Ohmic dissipation is very small in reality,
the ideal-MHD condition may be appropriate in the stellar interior.
In this case, we need not consider 
the differential rotation caused by Ohmic heating .
In the following discussion we assume the rigid rotation  in the form of
$\V = \vec{r} \times \Omega (t) \vec{e}_z$. 
The basic equations are
\begin{eqnarray} 
\label{torq.a-mhd1}
\rho {D\V \over Dt } 
&=& {1 \over c} \J \times \B - \grad \vec{\Pi} - \grad \Phi_g ,\\
\label{torq.a-mhd2}
\E + {1 \over c} \V \times \B &=& 0 ,
\end{eqnarray}
where $\rho$ is the mass density,
$\Phi_g$ is the gravitational potential, and
$\vec{\Pi}$ is the stress tensor.
The magnetic field $\B$ and the associated current $\J$ 
include the components that are ``intrinsic'' to the star, i.e., those
determined from the star structure and evolution.
We write them in the from of the intrinsic term plus the magnetospheric term:
$\B = \B_0 + \B^\prime$ and
$\J = \J_0 + \J^\prime$.
The expansion is
\begin{equation}
\J \times \B =
  \J_0 \times \B_0 
+ \J^\prime \times \B_0 
+ \J_0 \times \B^\prime
+ \J^\prime \times \B^\prime .
\end{equation}
Note that the  the magnetospheric components
are much smaller than the intrinsic ones 
($|\B^\prime |/|\B_0 | \sim (R_*/R_L)^{3/2}$).
The first term $\J_0 \times \B_0 /c$ is not related to the spin-down.
The dominant braking force is due to the term
$\J^\prime \times \B_0 /c$.
For  the axisymmetric cases, 
this term is  the only force in the
azimuthal direction. 
We ignore the last term.

Because we have specified the velocity field in terms of $\Omega (t)$ 
and have introduced elasticity,
the set of equations (\ref{torq.a-mhd1}) and (\ref{torq.a-mhd2}) have a solution for
any given magnetospheric current $\J^\prime$.
Then we cannot find a unique solution
for the magnetospheric current distribution.
However, if the stress is very high, the material will have viscous motion and/or
cracks to reduce the elastic energy $E_{\rm el}$.
The spin-down torque is too small to break the latices.
Hence, 
minimization of $E_{\rm el}$ will be achieved only 
through tiny non-elastic
change in the material, and thus the current distribution will 
change to the state in which $E_{\rm el}$ is minimized.
We take the following steps: 
(i) solving (\ref{torq.a-mhd1}) for a trial $\J^\prime$,
(ii) calculating $E_{\rm el}$ from the obtained displacement $u_\varphi$,
(iii)  varying $\J^\prime$ and then recalculating $E_{\rm el}$,
and 
(iv) finding $\J^\prime$ in  the searched parameter space so as to minimize $E_{\rm el}$.

For simplicity, we assume axisymmetry about and  uniform background magnetic fields
along the rotation axis, $\B_0 =B_0 \vec{e}_z$.
The star is assumed to be a uniform elastic body.
The azimuthal component of the equation of motion  is
\begin{equation} \label{torq-a-inside1}
\DDp{u_\varphi}{r} + {1 \over r} \Dp{u_\varphi}{r} - {u_\varphi \over r^2}
+ \DDp{u_\varphi}{z}
+ {F_\varphi - \rho \dot{\Omega} r \over \mu_{\rm s} }=0 ,
\end{equation}
where we use the cylindrical coordinates $(r, \varphi, z)$, 
for which the unit vectors are denoted by
$\vec{e}_r$, $\vec{e}_\varphi$ and $\vec{e}_z$,
$u_\varphi$ is the displacement in the azimuthal direction,
$F_\varphi = ( \J^\prime \times \B_0 ) \cdot \vec{e}_{\varphi}/c$
is the electromagnetic force  generated by the magnetospheric current,
and
$\mu_{\rm s}$ is the shear modulus which is uniform.
We may use the spherical polar coordinates $(R, \theta, \varphi)$ 
occasionally.

The magnetospheric current inside the star is purely poloidal and
is represented by, with the use of the current stream function
$I = r B^\prime_\varphi c /2$, 
\begin{equation}
\J^\prime_p = - {c \over 4 \pi r } \vec{e}_\varphi \times \grad (r B^\prime_\varphi)
 = - {1 \over 2 \pi r } \vec{e}_\varphi \times \grad I .
\end{equation}
It is notable that the contours of $I$ give the stream lines of the
current.
The value of $I$ on the star surface is denoted by $I_* (s)$,
which is a function of the arc length from the pole 
along  the meridian $s=R_* \theta$. 
The function $I_* (s)$ is treated as the boundary condition of the present model,
connecting the star and  magnetosphere.
Note that $I_*(s)$ gives the net current running through the surface
area  with an angular range of $0 \leq \theta \leq s/R_*$.

The azimuthal component of the electromagnetic force is
\begin{equation} 
\label{torq-a-emf}
F_{\varphi} = 
{1 \over c} (\J^\prime_p \times \B_0 )\cdot \vec{e}_\varphi
=  {1 \over 2 \pi c r}  \B_0 \cdot \grad I.
\end{equation}
The net torque on the star is straightforwardly obtained 
with surface integration to be 

\begin{equation}
\label{torq-a-torq}
N 
= \int_{\taiseki_*} r F_\varphi dV
= {1 \over 2 \pi c } \int_{\Surf_*} I_* (s)  \ \B_0 \cdot d \A. 
\end{equation}
Once $I_*(s)$  has been given as a boundary condition, 
$\dot{\Omega} = N/\Im$ is determined.

The shear stress  is  given by
\begin{eqnarray}
\Pi_{z\theta} &=& - \mu_{\rm s} \Dp{u_\varphi }{z} , \\
\Pi_{r\theta} &=& - \mu_{\rm s} \left( \Dp{u_\varphi }{r} - {u_\varphi  \over r} \right) .
\end{eqnarray}
The elastic energy is
\begin{equation}
E_{el} = \int_{\taiseki_*} { \Pi_{z\theta}^2 + \Pi_{r \theta}^2 \over \mu_{\rm s} } dV .
\end{equation}
If $I$ is specified, 
(\ref{torq-a-inside1}) is easily solved numerically, and
the elastic energy is obtained.

\begin{figure*}
\begin{center}
\includegraphics[width=6cm,page=2]{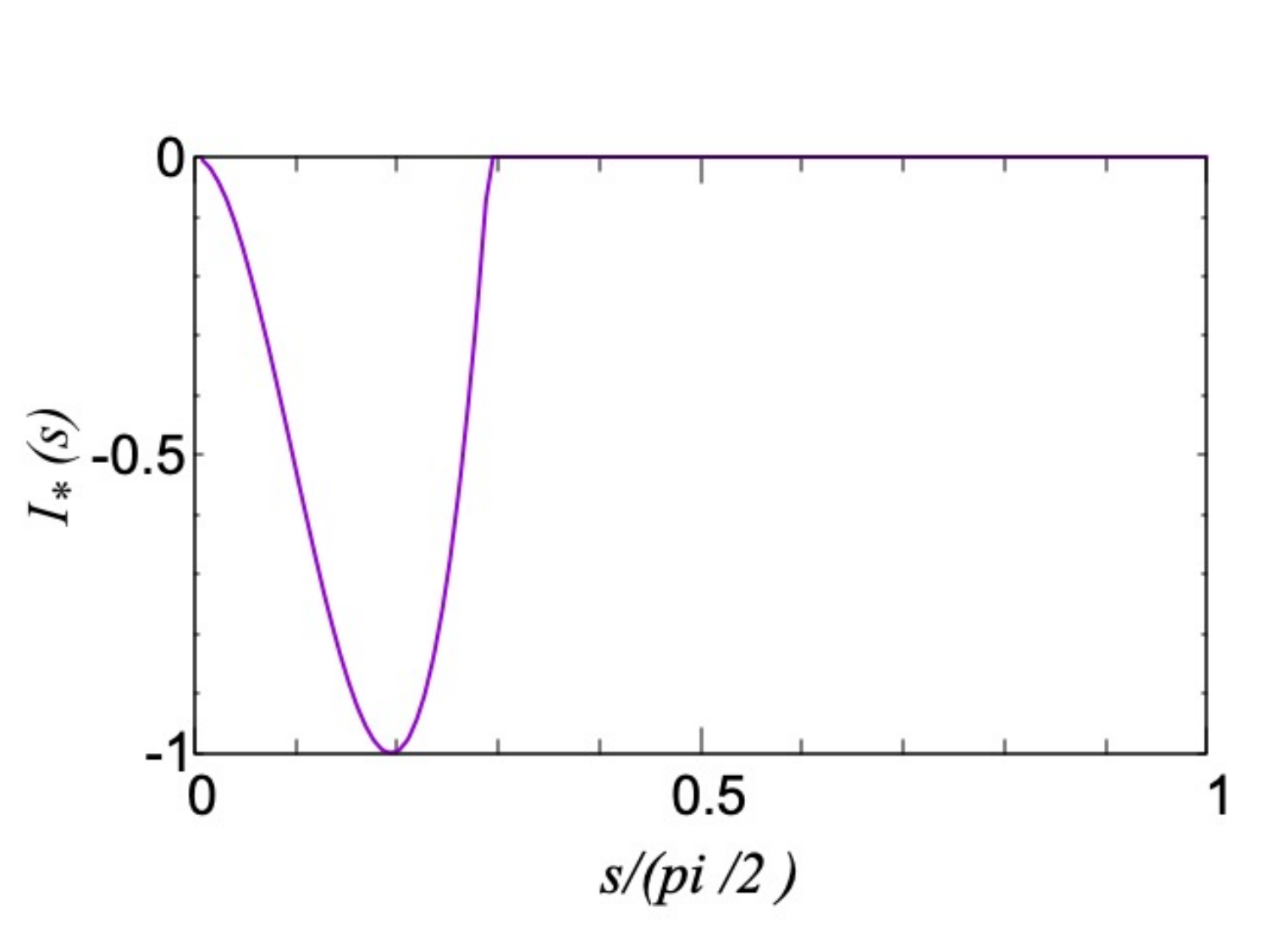}
\includegraphics[width=6cm,page=3]{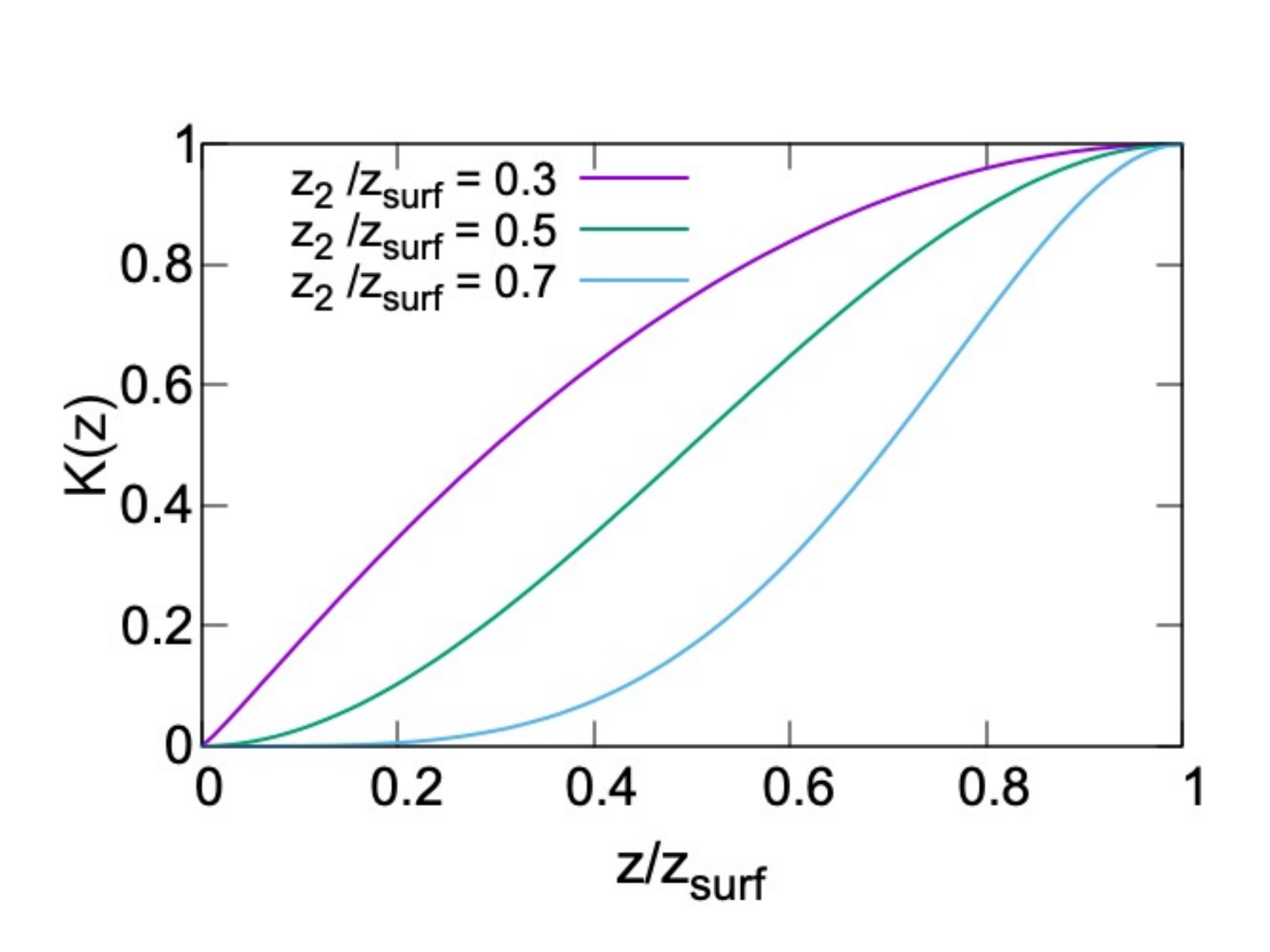}
\caption{\label{torq.b-fig9} (Left panel) 
Current function $I_* (s)$ (Equation (52)) 
for the surface boundary, where
$s$ and $I_*$ are in units of 
$R_*$ and $I_0$, respectively.
(Right panel) 
Distribution $K(z)$ (Equation (53)right panel)  as a function of $z$.
} 
\end{center}
\end{figure*}

\begin{figure*}
\begin{center}
\raisebox{5cm}{(a)}
\includegraphics[width=75mm,page=4]{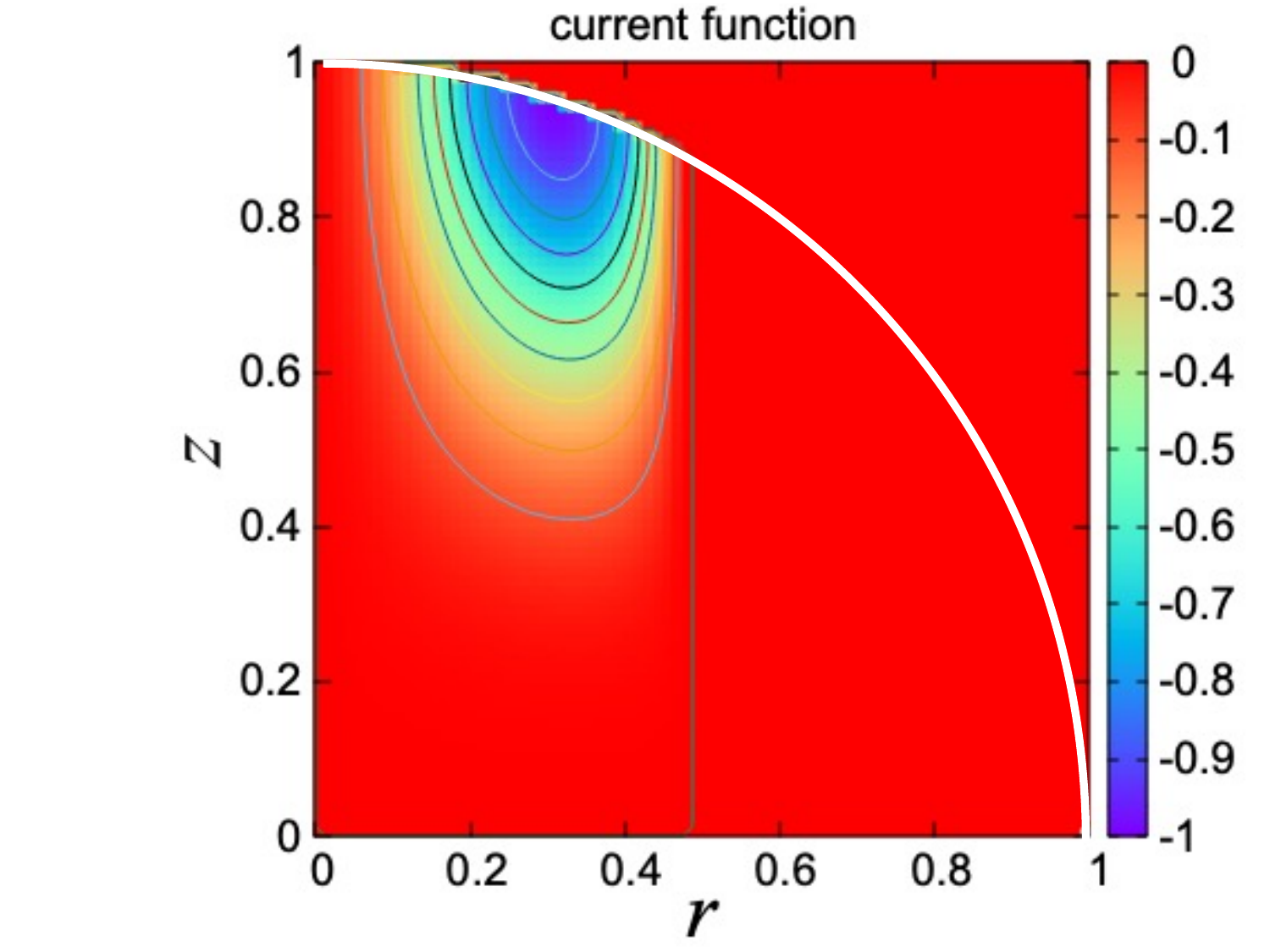}
\raisebox{5cm}{(b)}
\includegraphics[width=75mm,page=5]{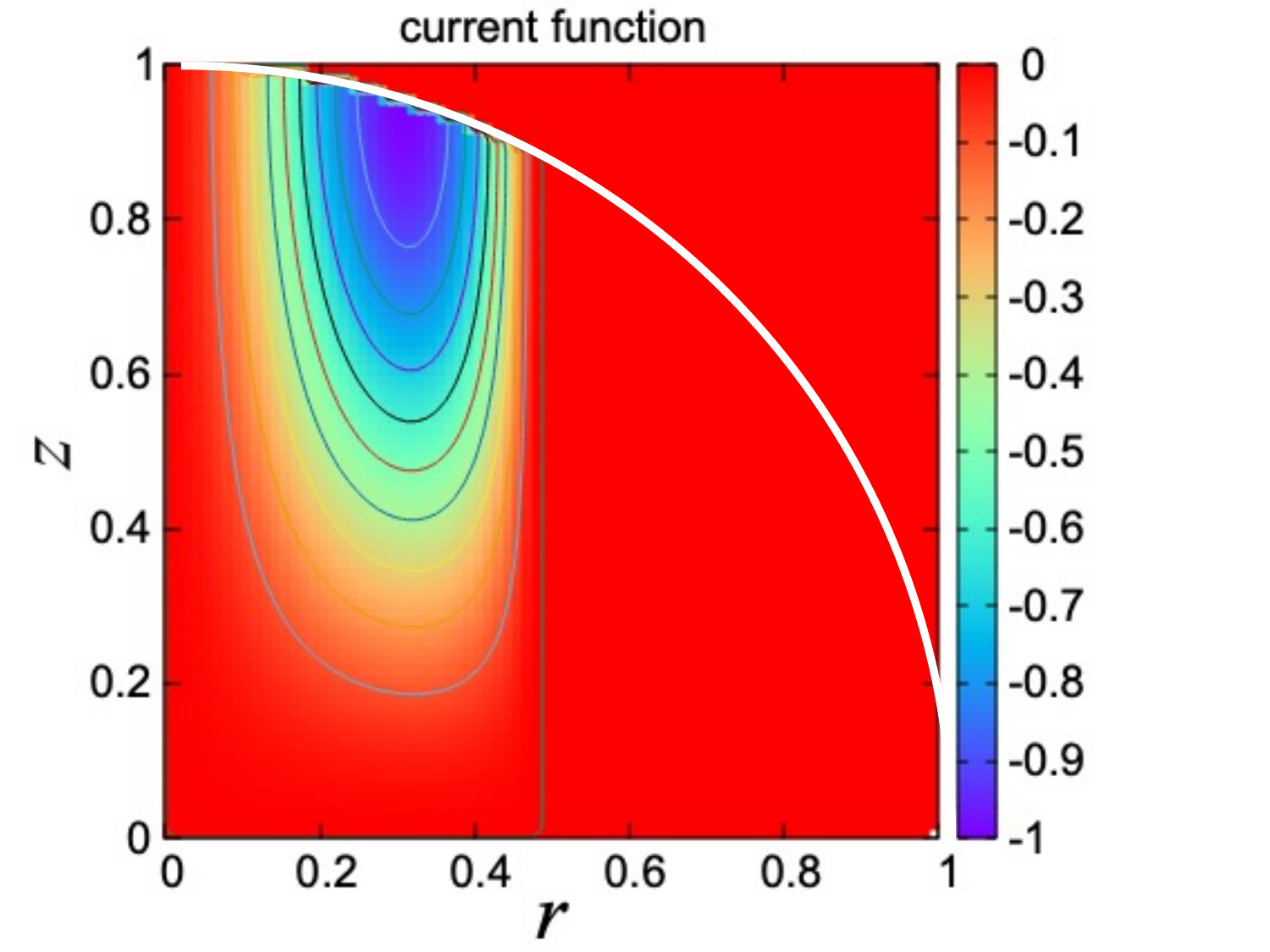} \\
\raisebox{5cm}{(c)}
\includegraphics[width=75mm,page=6]{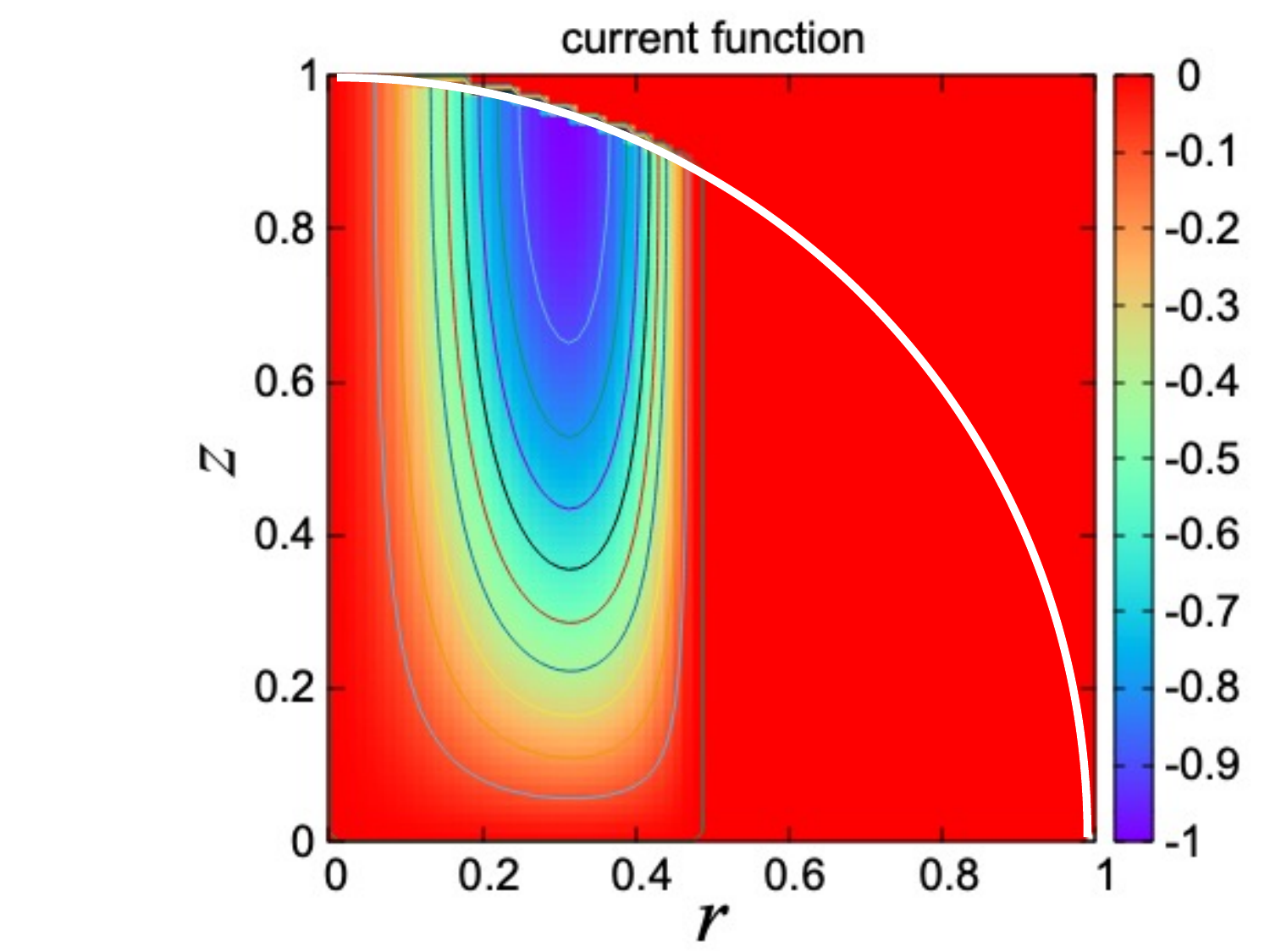}
\raisebox{5cm}{(d)}
\includegraphics[width=75mm,page=7]{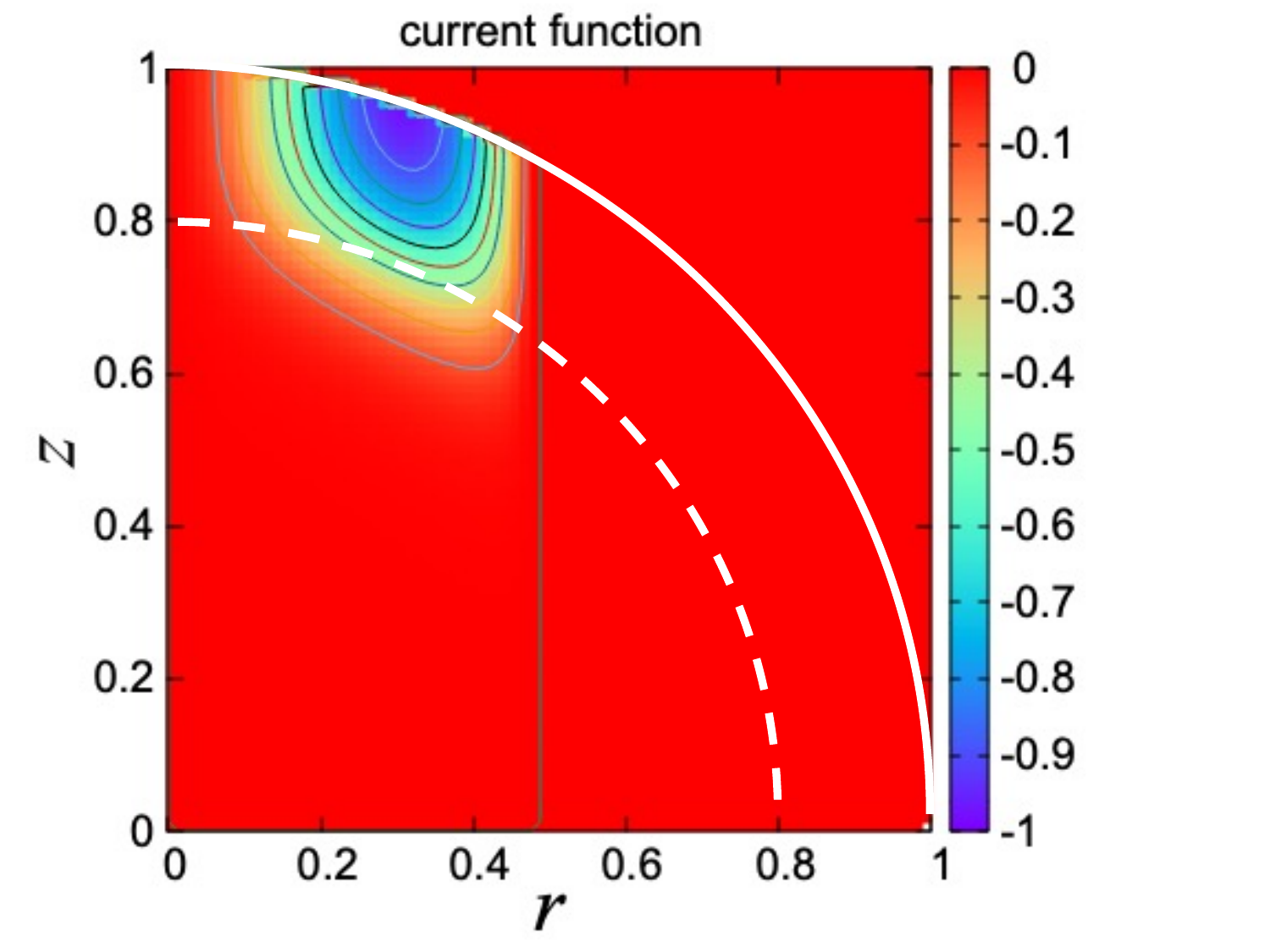}
\caption{ \label{torq.b-fig10} 
Contour maps of the current function $I(r,z)$ 
(Equation (\ref{torq-I}) ) with
(a) $z_2=0.7 z_{\rm surf} $, 
(b) $z_2=0.5 z_{\rm surf} $, 
and
(c) $z_2=0.3 z_{\rm surf} $, without a diminishing factor, and 
(d) $z_2 =0.5 z_{\rm surf} $ with a diminishing factor 
$K_{\rm cut} (R)$ (Equation(\ref{torq-K})) applied with $R_1 = 0.8 R_*$ and  
$\Delta R=0.05 R_*$. 
The coordinates are in units of $R_*$.
}
\end{center}
\end{figure*}

We need to provide some trial current function. 
In this study, we employ for it 
simple cubic forms with a few parameters,
which are determined so  that  $E_{\rm el}$  is minimized.
For $I_*(s)$,
a simple analytic function
\begin{equation}
I_*(s) = \left\{ \begin{array}{lc} 
\displaystyle
I_0 \ {27 \over 4} {s^2 (s-s_{\rm pc} ) \over s_{\rm pc}^3 } &
\mbox{if } s< s_{\rm pc} \\
0 &
\mbox{if } s> s_{\rm pc} 
\end{array}
\right.
\end{equation}
is assumed, where
$s_{\rm pc} =R_* \theta_{\rm pc} $ with the polar cap co-latitude
$\theta_{\rm pc}=\sqrt{R_* \Omega_* /c}$ 
(Fig.~\ref{torq.b-fig9}), and $I_0$ is the peak value, meaning
the total circulating current.
In the following numerical illustration , we use a slightly large value
of $\theta_{\rm pc} = (R_L / R_*)^{-1/2} = 0.2913$ 
(corresponding to a 1-msec pulsar) to reduce numerical load.
The current function within the star $I (r,z)$ describes how current runs
inside the star. 
We additionally introduce a simple cubic function $K$ with the parameter $z_2$
above which one half of the current $I_0$ closes:
\begin{equation}
{K} (\zeta) = \zeta^2 (3-2\zeta)
\end{equation}
with
\begin{eqnarray}
\zeta(z) &=& \left( z \over z_{\rm surf} \right)^{1/\alpha} , \\
\alpha &=& - { \log_2 z_2 } , 
\end{eqnarray}
where $z_{\rm surf}=\sqrt{R_*^2 - r^2}$ 
is the $z$-value on the surface for a given $r$
(right panel of Fig.~\ref{torq.b-fig9}).
Note that 
${K} (0) = 0$,
${K} (1/2) = 1/2$,
${K} (1) = 1$,
and 
${K} (\zeta(z=z_2)) = 1/2$.
Although how deep the magnetospheric current penetrates into the star is parameterized by $z_2$,
one may  need to restrict the current in a layer $R_1 < R < R_*$, say the crust.
If this is the case, we use a diminishing factor of the Fermi-Dirac type:
\begin{equation} \label{torq-K}
K_{\rm cut} (R) = { 1 \over \exp [ (R_1 - R)/\Delta R ] +1 } ,
\end{equation}
where $\Delta R$ is the thickness for the transition. 
If not, 
 $K_{\rm cut} = 1$.
Incorporating this factor $K_{\rm cut}$, we have
\begin{equation} \label{torq-I}
I(r,z) = K_{\rm cut} (R) \ K (\zeta(z)) \ I_*(s(r)),
\end{equation}
where $R=\sqrt{r^2 + z^2}$, and
$s = R_* \arcsin (r/R_*)$.
Some examples of $I(r,z)$ are shown in Fig.~\ref{torq.b-fig10}. 
As $z_2$ is decreased  (see  panels (a), (b) and (c), 
of Fig.~\ref{torq.b-fig10} in this order),
the current runs deeper along the field lines. If the diminishing factor is
applied, the current is restricted above $R_1$ 
(panel (d)).

\begin{figure*}
\begin{center}
\raisebox{5cm}{(a)}
\includegraphics[width=79mm,page=8]{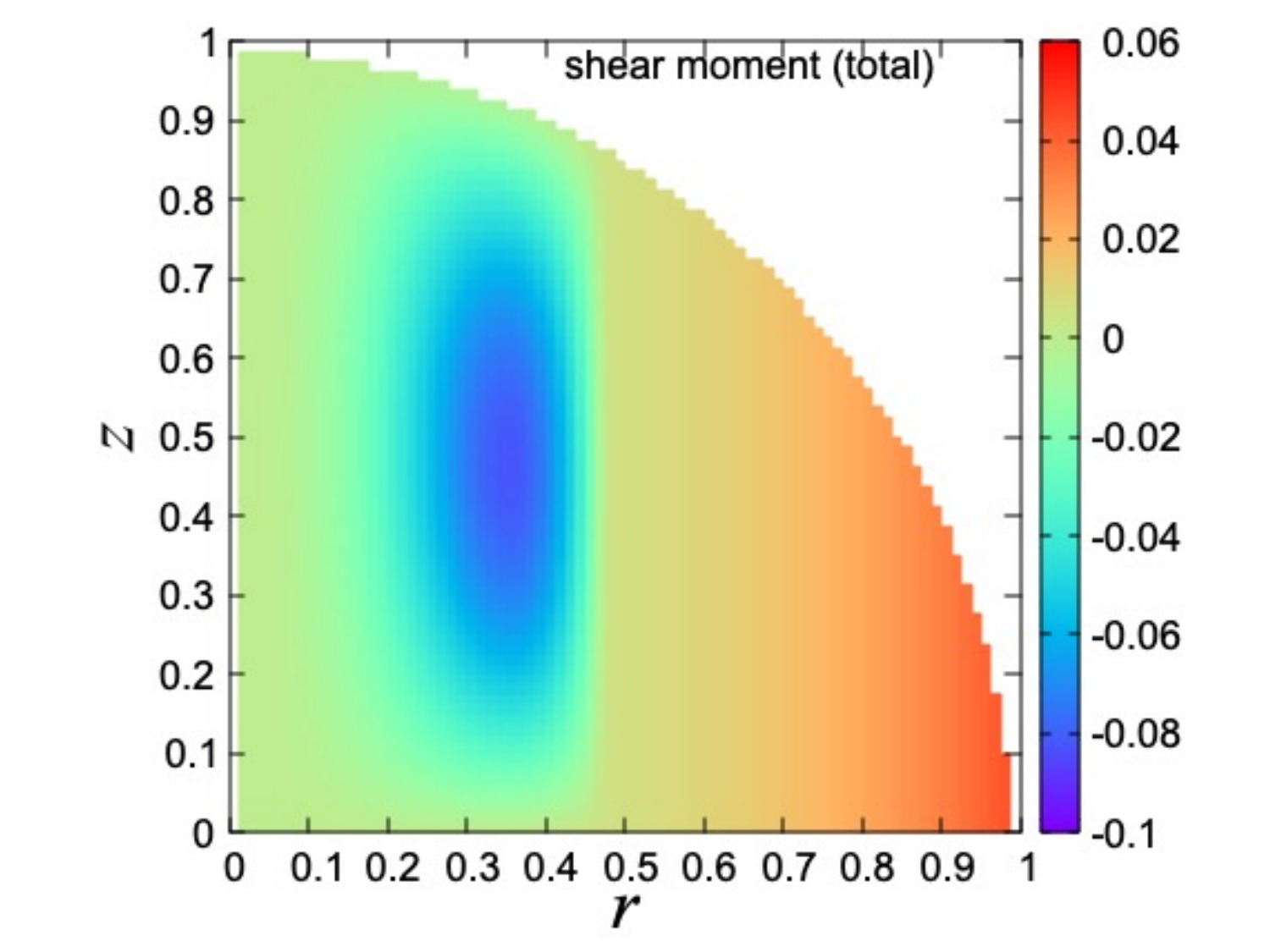}
\raisebox{5cm}{(b)}
\includegraphics[width=79mm,page=9]{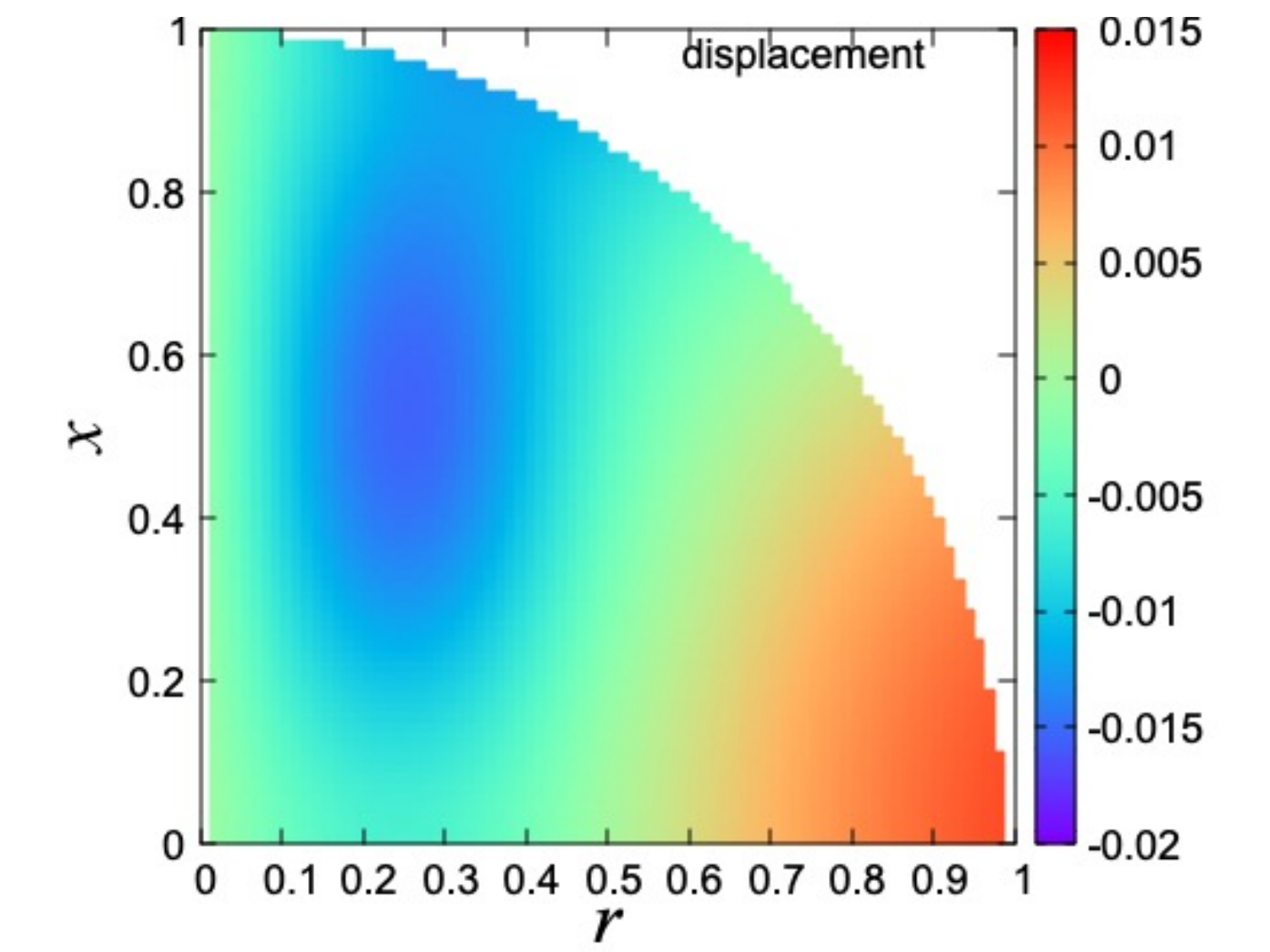} \\
\raisebox{5cm}{(c)}
\includegraphics[width=79mm,page=10]{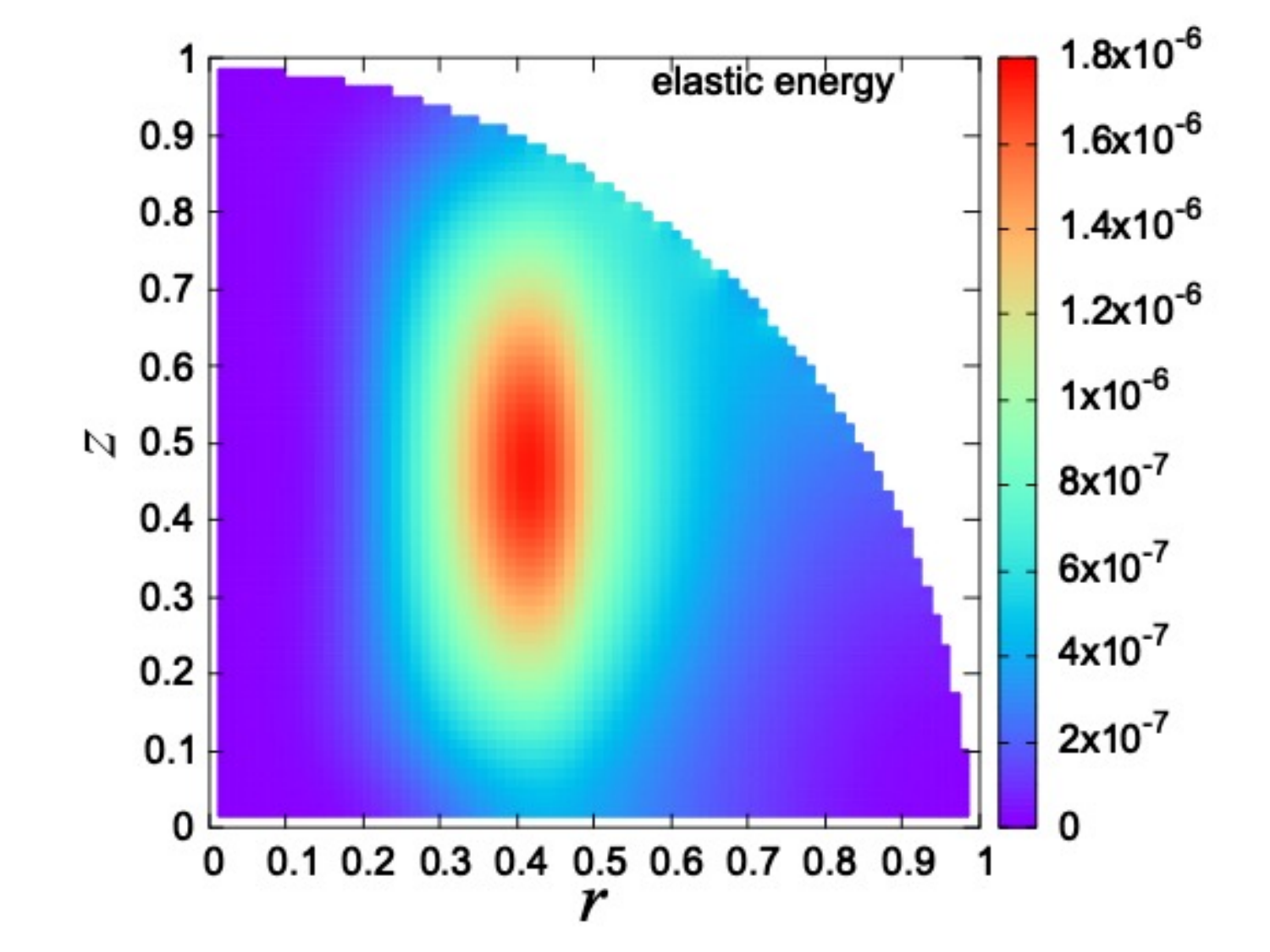}
\caption{ \label{torq.b-fig14} 
(a)  Distribution of force  represented by the shear moment,
(b)  obtained displacement $u_\varphi$,
and
(c)  elastic energy density
for the case of $z_2=0.5z_{\rm surf}$
(Fig.~\ref{torq.b-fig10}(b)).
The  shear moment, displacement  and $E_{\rm el}$ are 
in units of $F_0 R_*^2$,
$F_0 R_*^2 / \mu_{\rm s}$,
and
$\dot{E}_0 \Omega_*^{-1}$,respectively.
}
\end{center}
\end{figure*}

For the given current, we calculate $F_\varphi$ and $\dot{\Omega}$
from  (\ref{torq-a-emf}) and (\ref{torq-a-torq}).
Fig.~\ref{torq.b-fig14}(a)
shows the shear moment $(F_\varphi - \rho \dot{\Omega}r ) r^2$
for the case of $z_2 = 0.5 z_{\rm surf} $. 
It shows  that 
the inertial force in $+\varphi$ direction dominates 
in the equatorial region, 
whereas
the electromagnetic braking force in $-\varphi$ direction
does in the region
$0 \leq r \leq R_* \sin \theta_{\rm pc}$, which is the polar cap magnetic flux.

With the specified $F_\varphi$ and $\dot{\Omega}$,
we solve (\ref{torq-a-inside1}) numerically to obtain the 
displacement $u_\varphi$.
The obtained displacement is shows in
Fig.~\ref{torq.b-fig14}(b).
As is expected, 
the inner ($r \la 0.5$) and outer ( $r \ga 0.5$) regions are
deformed in $+\varphi$ and $-\varphi$ directions, respectively.
In the boundary region between the two, 
the elastic energy  is large  as shown in
Fig.~\ref{torq.b-fig14}(c).
The numerical calculation is done in the non-dimensional form,
where all the formulae are the same except $c=1$.
The force density, displacement and $E_{\rm el}$ are respectively
in units of $F_0=B_0 I_0 /c R_*^2$,
$F_0 R_*^2 / \mu_{\rm s}$,
and
$\dot{E}_0 \Omega_*^{-1}$,
where 
the unit of the power 
is $\dot{E}_0 = (\mu_{\rm m}^2 \Omega_*^4 /c^3 ) (2R_L/R_*)$,
and $\mu_{\rm m}$ is the magnetic moment of the star.

\begin{figure*}
\begin{center}
\raisebox{5cm}{(a)}
\includegraphics[width=7cm,page=11]{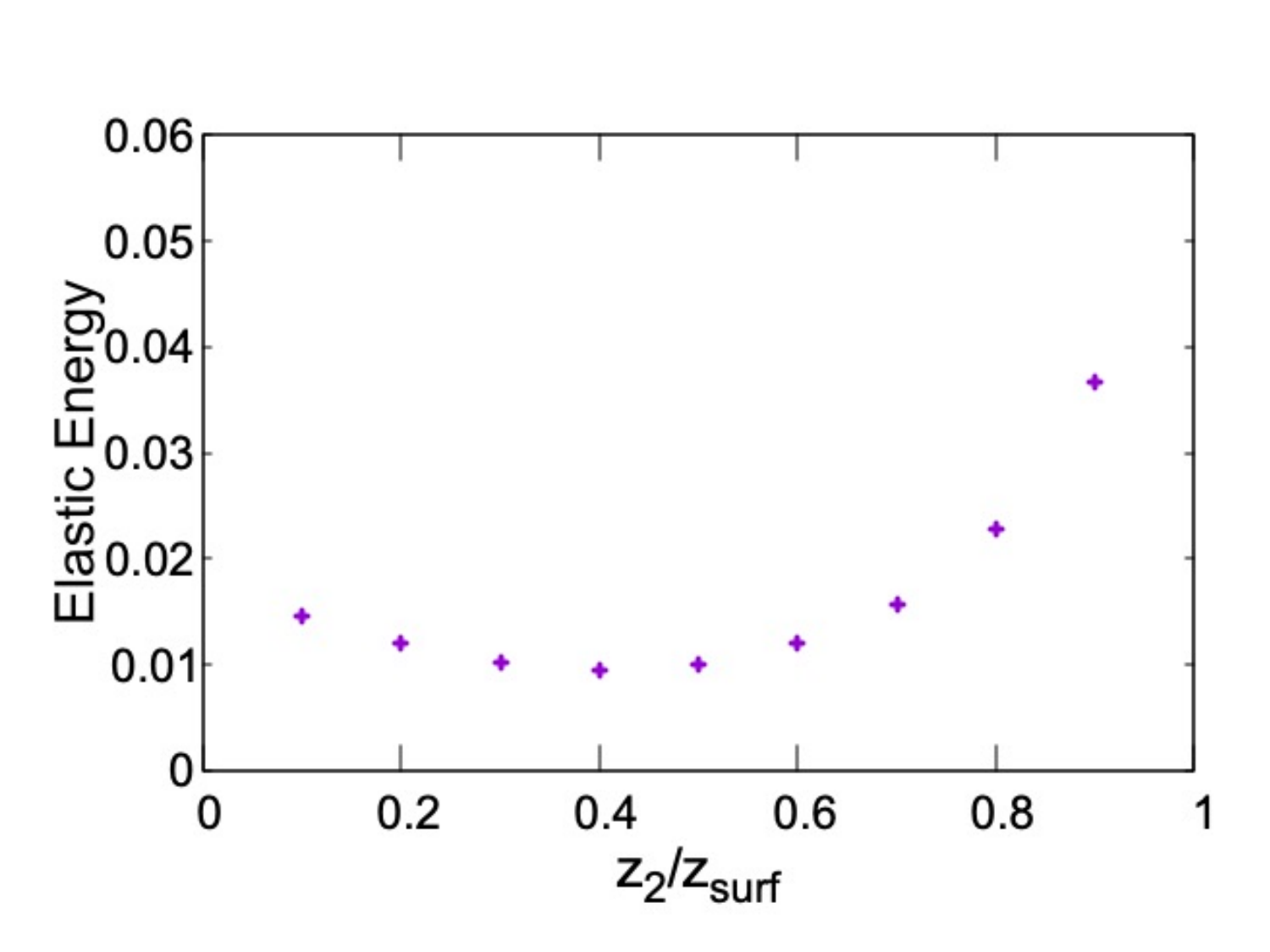}
\raisebox{5cm}{(b)}
\includegraphics[width=7cm,page=12]{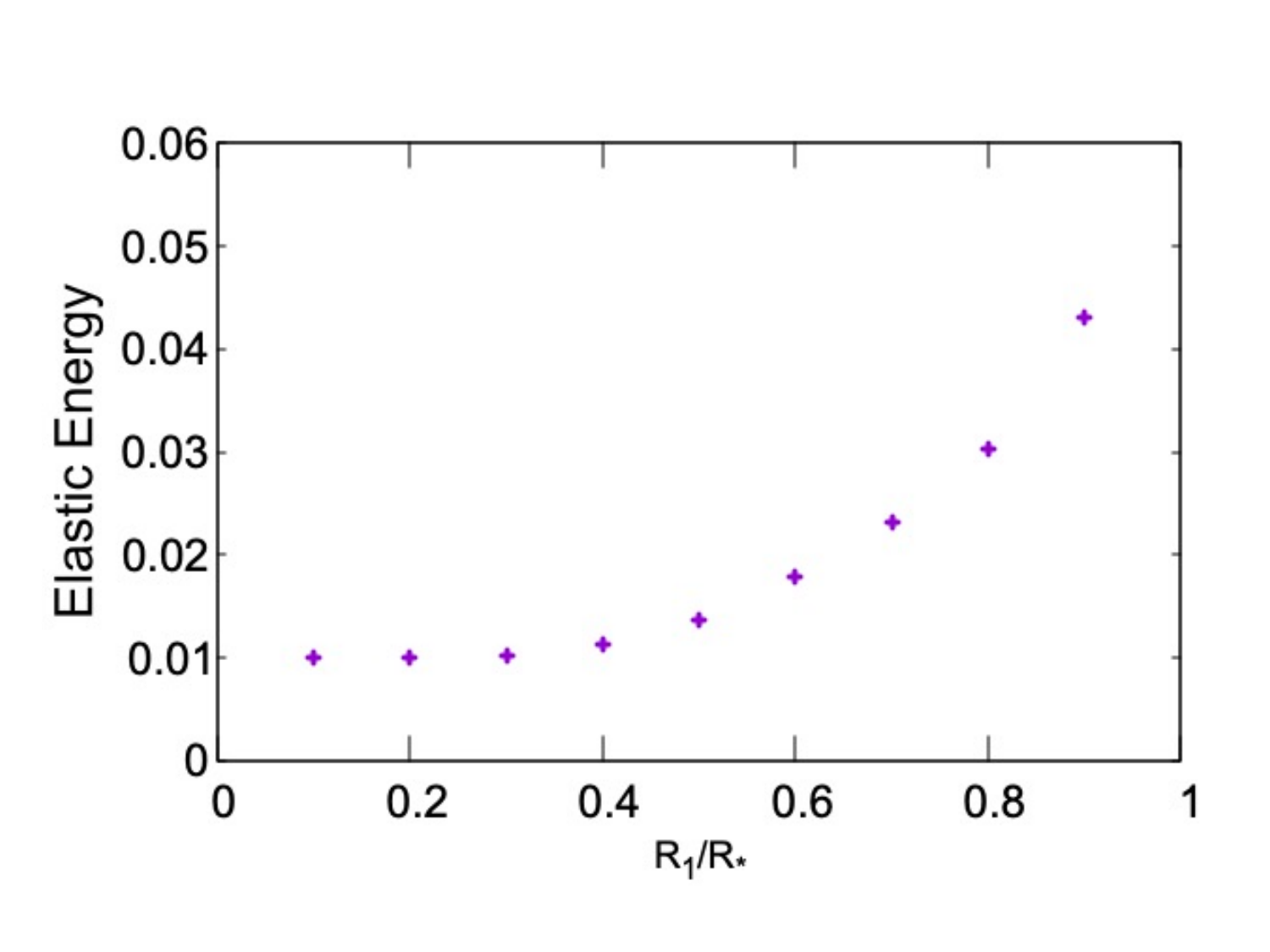} \\
\caption{ \label{torq.a-fig13} 
Total elastic energy as a function of (a) $z_2 / z_{\rm surf} $ with $K_{\rm cut}=1$
and (b) $R_1$ with the fixed $z_2 = 0.5 z_{\rm surf}$.
}
\end{center}
\end{figure*}

We then obtain a series of solutions  for 
different $z_2$ and
find that the elastic energy is minimized at $z_2 \sim 0.4 z_{\rm surf} $
as shown in 
Fig.~\ref{torq.a-fig13}(a).
This result is plausible, i.e.,
the elastic energy is decreased if the braking force is distributed
over the whole neutron star  to diminish the inertial force.
Conversely, if the braking force is localized in a small region
near the surface region ($z_2 \sim 1$) or
the core region ($z_2 \sim 0$), 
then the elastic energy is increased.
If the current is restricted in a layer above $R_1$,
the  thinner the current is restricted, 
the larger the elastic energy is
(Fig.~\ref{torq.a-fig13}(b)).

In addition to this,
if the current penetrates out of the polar cap magnetic flux tube
($r > R_* \sin \theta_{\rm pc}$),
the electromagnetic force  works in the opposite direction  to the braking force
($+\varphi$ direction),
and the elastic energy increases.
As a result, 
the magnetospheric current inside the star  is restricted in the
polar cap magnetic flux tube.
Thus,
the configuration of $\B_0$, especially that of the polar cap magnetic 
flux plays essential role to relax the elastic energy of the star.
This  point is discussed in {\S}~\ref{discussB}.

\section{Discussion} \label{discuss}

\subsection{Outer magnetosphere} \label{discussA}

The angular momentum extracted from the neutron star 
must be the rotation power divided by the angular velocity, 
$\dot{L}= \dot{E} /\Omega_*$.
However,
the polar cap accelerator emits a part of the rotational energy
but emits little angular momentum.
We have shown in \S~\ref{magsec} that
the inefficient loss of angular momentum is compensated by efficient loss of
angular momentum  in the outer magnetosphere beyond the light cylinder.
The surface $\Surf$ in our simple model
(Fig.~\ref{torq-fig34}) is effectively a slow rotator.
Since the efficiency depends on 
$\Omega_{out}^{-1} =\dot{L}_{\rm out} /\dot{E}_{\rm out}$,
a slower rotation than $\Omega_*$ (sub-rotation)  plays a key role.
This effect   is  more significant for older pulsars, for which 
the polar cap accelerator carries off an  major fraction of
the rotation power.

\begin{figure}
\begin{center}
\includegraphics[width=7cm,page=13]{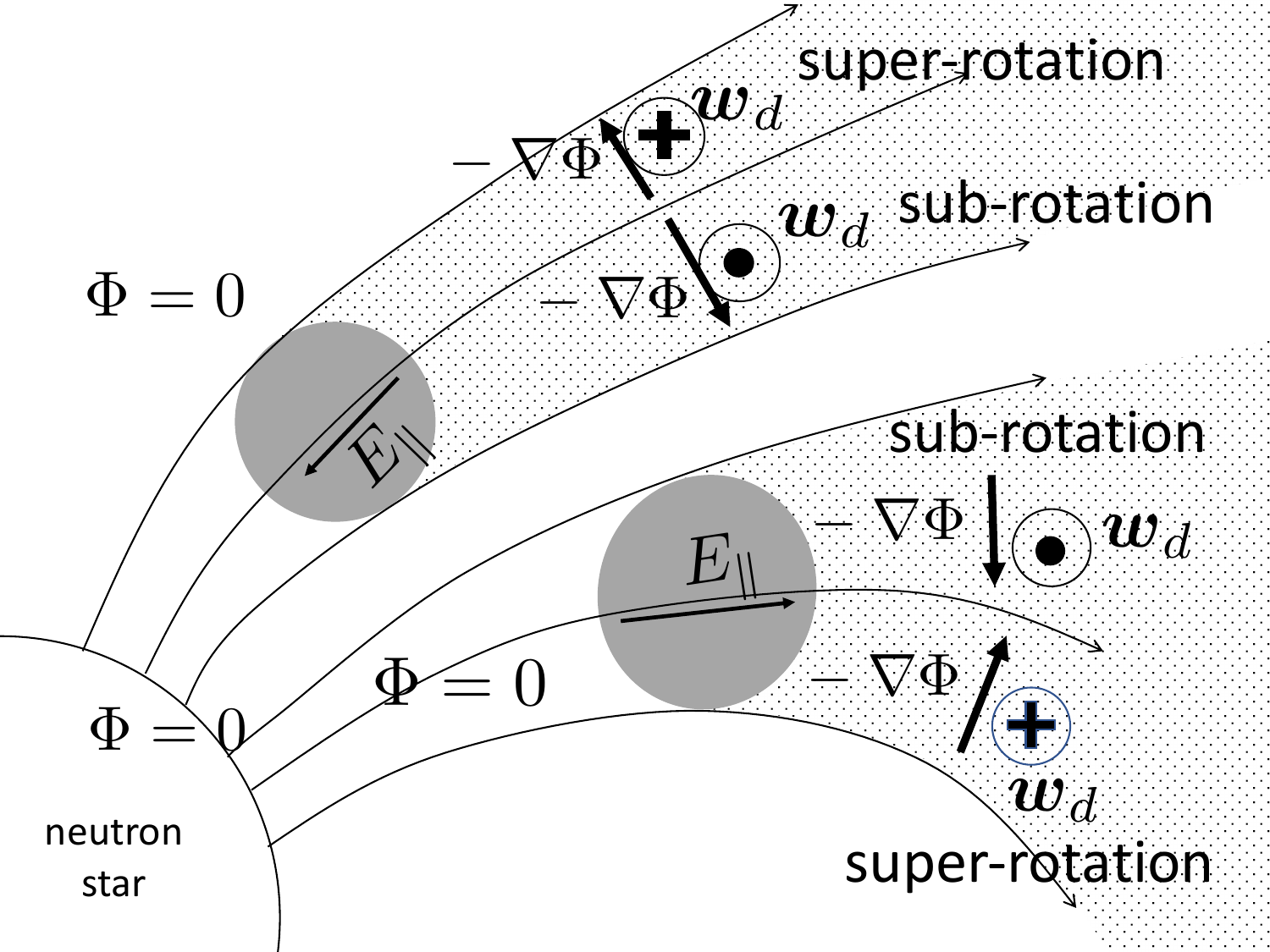}
\caption{ \label{torq.b-fig23} 
Field-aligned particle acceleration results in  sub-rotation and
super-rotation regions along field lines piercing  field-aligned accelerators.
The sub-rotating magnetic fluxes contribute to the efficient loss of angular momentum.
The non-corotational potential $\Phi$ is zero everywhere except for the
shaded regions. 
It is notable that the dotted regions satisfy the ideal-MHD condition but has
the gradient of $\Phi$.
}
\end{center}
\end{figure}

Let us consider what actually happens in the outer magnetosphere.
The outer gap around the null surface  is located within the light cylinder
\citep{1973NPhS..246....6H} 
and therefore  cannot account for the expected loss of the angular momentum. 
The polar cap accelerator and  outer gap 
causes gradients of the
non-corotational potential $\Phi$.  As a result, the regions outside these
accelerators show the sub-rotation and
the super-rotation 
by the non-corotational drift motion according to (\ref{torq.a-drift}).
Fig.~\ref{torq.b-fig23} illustrates the drift motion in the case of
a nearly aligned rotator. 
We argue that
the energy is conveyed off efficiently 
along the sub-rotating open field lines
to compensate the insufficient loss of angular momentum.

The emission beyond the light cylinder might be essential.
\citet{2003ApJ...598..446U}  
showed that the ideal-MHD condition breaks down 
beyond the Y-point, the open-close boundary on the light cylinder.
He showed that a wedge shaped region beyond the Y-point 
 is electric field dominant,  $|E| >|B|$, and
the force-free approximation is no longer applicable there.
This region indeed appears in  particle simulations
\citep{{2011MNRAS.418..612W},{2012PASJ...64...43Y}}.
It has been suggested that the  location of the  Y-point is a free
parameter for the force-free model 
\citep{2006MNRAS.368.1055T}. 
This fact alone might indicate that the Y-point shift outward so that
the  lever arm becomes larger than the light cylinder radius.
However, as was suggested by \citet{1973NPhS..246....6H} 
and seen in 
Fig.~\ref{torq.b-fig23} as well as in the numerical simulation 
\citep{{2011MNRAS.418..612W},{2012PASJ...64...43Y}},
the region  around the
open-close boundary is super-corotation,  and hence
the Y-point shifts inwards.
The particle acceleration and radiation in the vicinity of the Y-point 
will not contribute to the efficient loss of the angular momentum,
but the current sheet beyond the light cylinder will do.
Some indication may be found in numerical simulations 
(e.g., \citep{2018ApJ...858...81B}).

\begin{figure}
\begin{center}
\includegraphics[width=7cm,page=14]{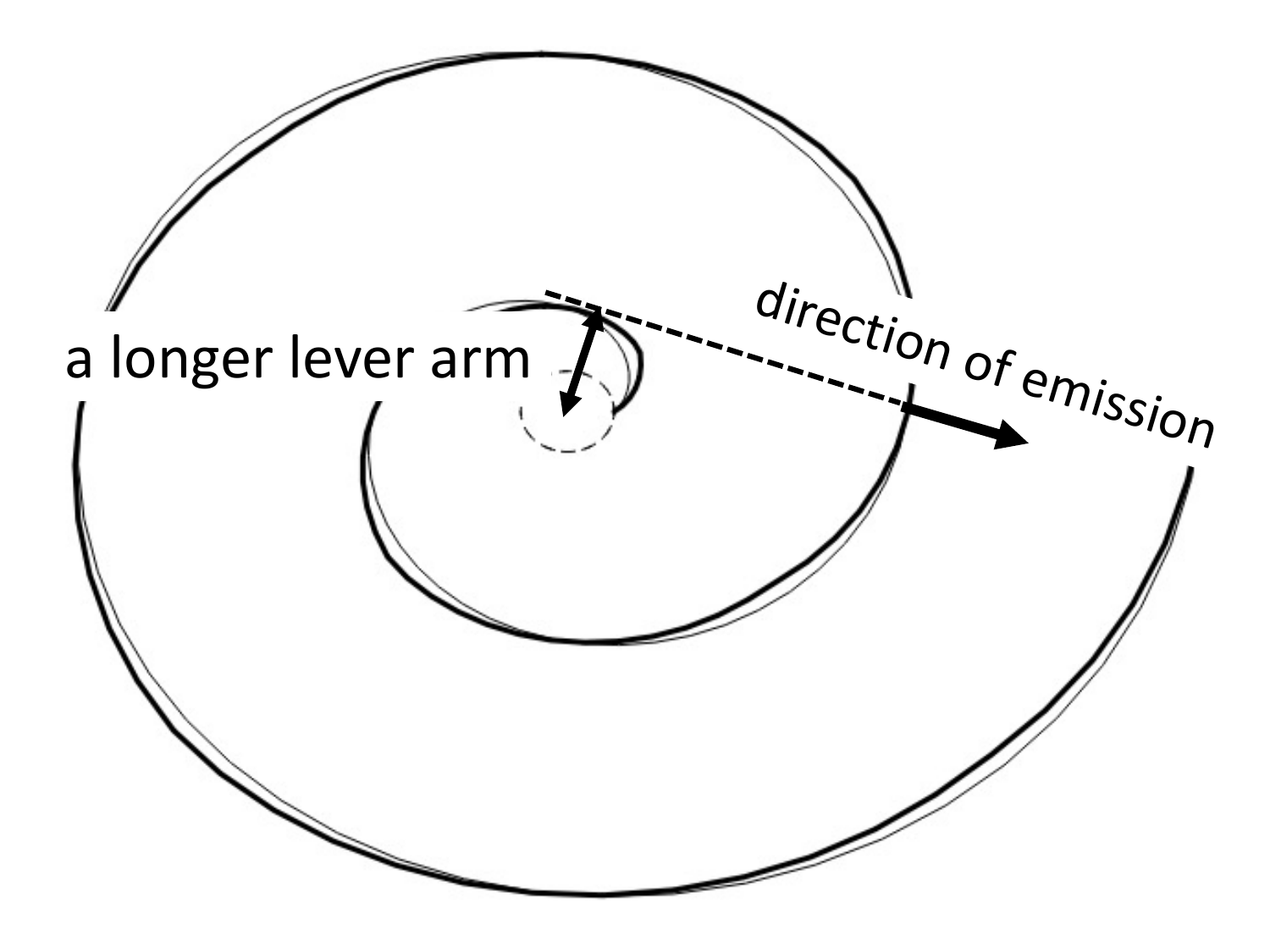}
\includegraphics[width=7cm,page=15]{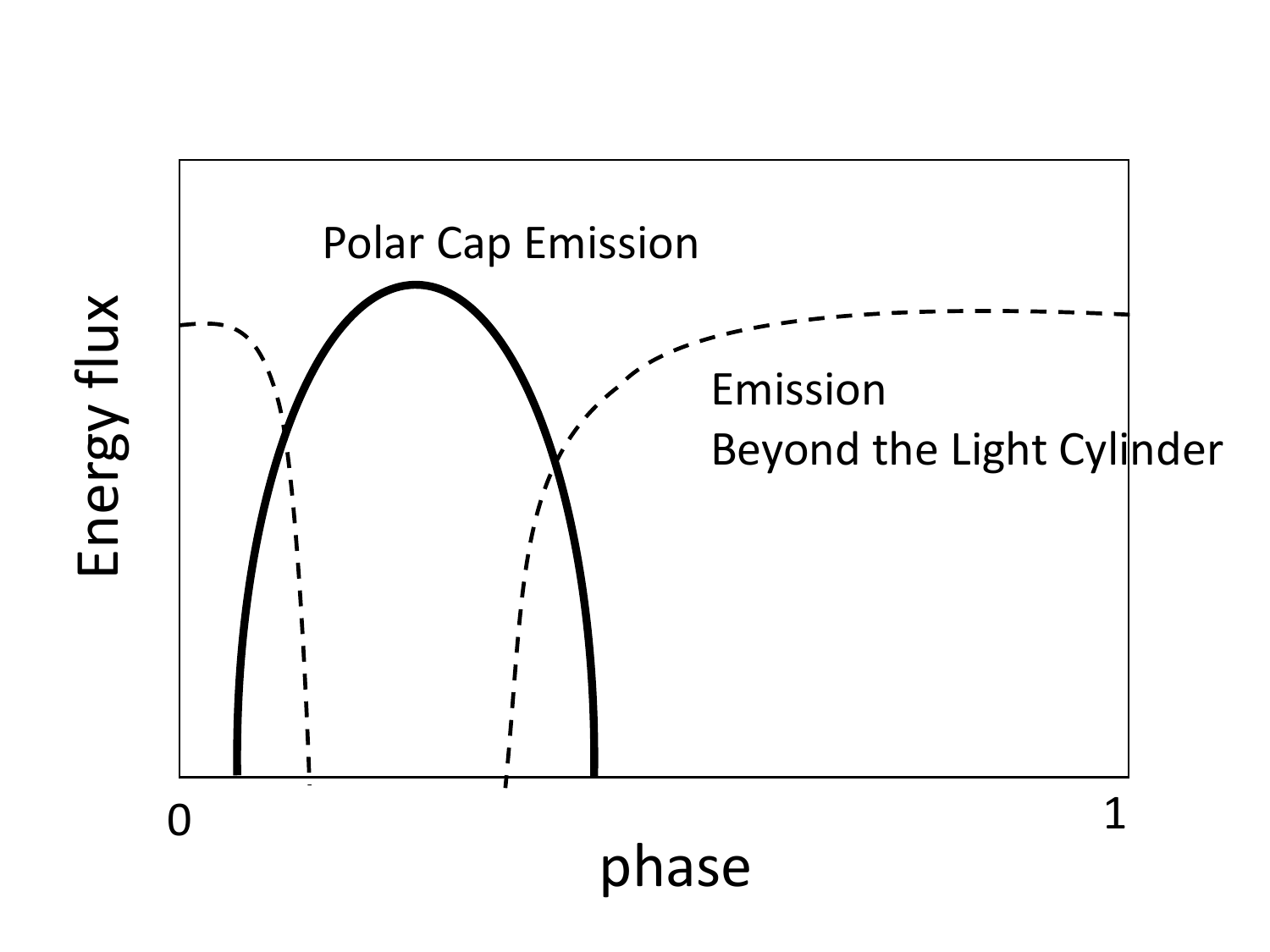}
\caption{ \label{torq.b-fig24} 
(Left panel)  Face-on view of the field lines in the wind region. 
The larger the pitch angle of  spiralling magnetic fields is, 
the larger the lever arm is,  and the more efficient
the angular momentum loss is.
The average pitch angle must be $c/\Omega_* r$ (the
thin spiral curve). In some phase angles  
as long as the system maintains the steady state.  
the pitch angle must be larger than the average value
to carry a larger amount of  angular momentum. 
(Right panel)
a large amount of energy may reside  out of the phase in which the
polar caps emit.
}
\end{center}
\end{figure}

The compensation is not always  associated 
with particle acceleration and  subsequent radiation. 
It can be made  in a form of a Poynting dominant wind, where
the required condition would be $|\vec{r} \times \vec{S} | / |\vec{S} | > R_L$.
Fig.~\ref{torq.b-fig24} left panel
shows the  face-on view of the wind region with spiralling magnetic field lines.
The pitch angle of the field lines is $\psi = c/ \Omega_* r \ll 1$ at large distances
on average.
This is  due to the fact that 
every quantity should have the pattern velocity $\Omega_*$
in steady state.
However, some part can have a larger pitch angle than the average, 
and if so, the Poynting vector has a larger lever arm 
than the light-cylinder radius.
Therefore, if
the wind energy is concentrated in the part with such a large pitch angle, 
the Poynting dominant wind conveys a larger amount of the angular momentum
than 
the value  of the wind power divided by $\Omega_*$.
In this case,
the energy flux is not uniform
with respect to the azimuthal angle, which implies
that the pulsar wind has amplitude modulation.
It is plausible that 
emission beyond the light cylinder may be observed out of the phase of
the polar cap emission as shown in the right panel of Fig.~\ref{torq.b-fig24}.

\subsection{Magnetospheric current inside the star} \label{discussB}

Ohmic dissipation of the magnetospheric current inside the star
converts the rotation energy into heat,
which is eventually radiated away.
No angular momentum
is lost in this process.
We have shown that
the Ohmic heating results in different angular velocities
between the surface and interior of the star.
Consequently the way 
the magnetospheric current inside the star closes
is coupled with the angular momentum transfer in the star and accordingly cannot
 be solved solely by Ohm's law, as erroneously
done in \citet{2019MNRAS.487.3333K}. 

In fact, the Ohmic heating is very small;
the heating rate is given by $H \approx (j^2/\sigma) V$, 
where $V \approx \varepsilon R_{\rm pc}^2 R_*$ is the volume and 
$\varepsilon$ is a numerical factor.
Using $j \approx I_0/R_{\rm pc}^2$ and $\dot{E} \approx I_0^2/c$, we have
\begin{equation}
H = \dot{E} \left( R_L \over R_* \right)^2 {\varepsilon \Omega \over \sigma} .
\end{equation}
Adopting a typical value $\sigma \sim 10^{25}$~s$^{-1}$
\citep{2008LRR....11...10C}, 
we have
$H /\dot{E} =  \exx{1.4}{-17} ( \varepsilon P / 1 \mbox{s})$,
where $P$ is the rotation period.
The lack of the angular momentum release
 accumulates
with time; e.g., 
in a month, $\Delta t \sim 10^6$~s, $H \Delta t / \Im \Omega^2 $
$= \Delta \Omega / \Omega \sim \exx{4}{-11} (\varepsilon  P / 1 \mbox{s})$.
This indicates that the insufficient loss cased by Ohmic heating 
contribute to some {kinds of timing irregularities, such as } glitches or some timing noises.
But they are small 
as compared with the observed limit \citep{2017A&A...608A.131F}.

We have given an illustrative example for the case  the ideal-MHD
condition holds (no Ohmic loss) and the star is an elastic solid body
in \S~\ref{currentdistribution}.
The current is determined  in such a way that the elastic energy is minimized. 
As demonstrated by the numerical solution, the elastic energy is
minimized when the current spreads over within the polar cap 
magnetic flux. 
In fact, if the elastic energy is high,
the situation is relaxed due to small non-elastic changes or viscous motions
and 
the current distribution is varied to reduce the elastic energy.
In contrast to the Ohmic law, where the current tries to find a short cut
\citep{{2015SSRv..191..207B}, {2019MNRAS.487.3333K}},
the actual current distribution is determined in such a way that
electromotive force is gained as efficiently as possible with least stress, and thus
 the current spreads over the stellar volume.

\begin{figure}
\begin{center}
\includegraphics[width=7cm,page=16]{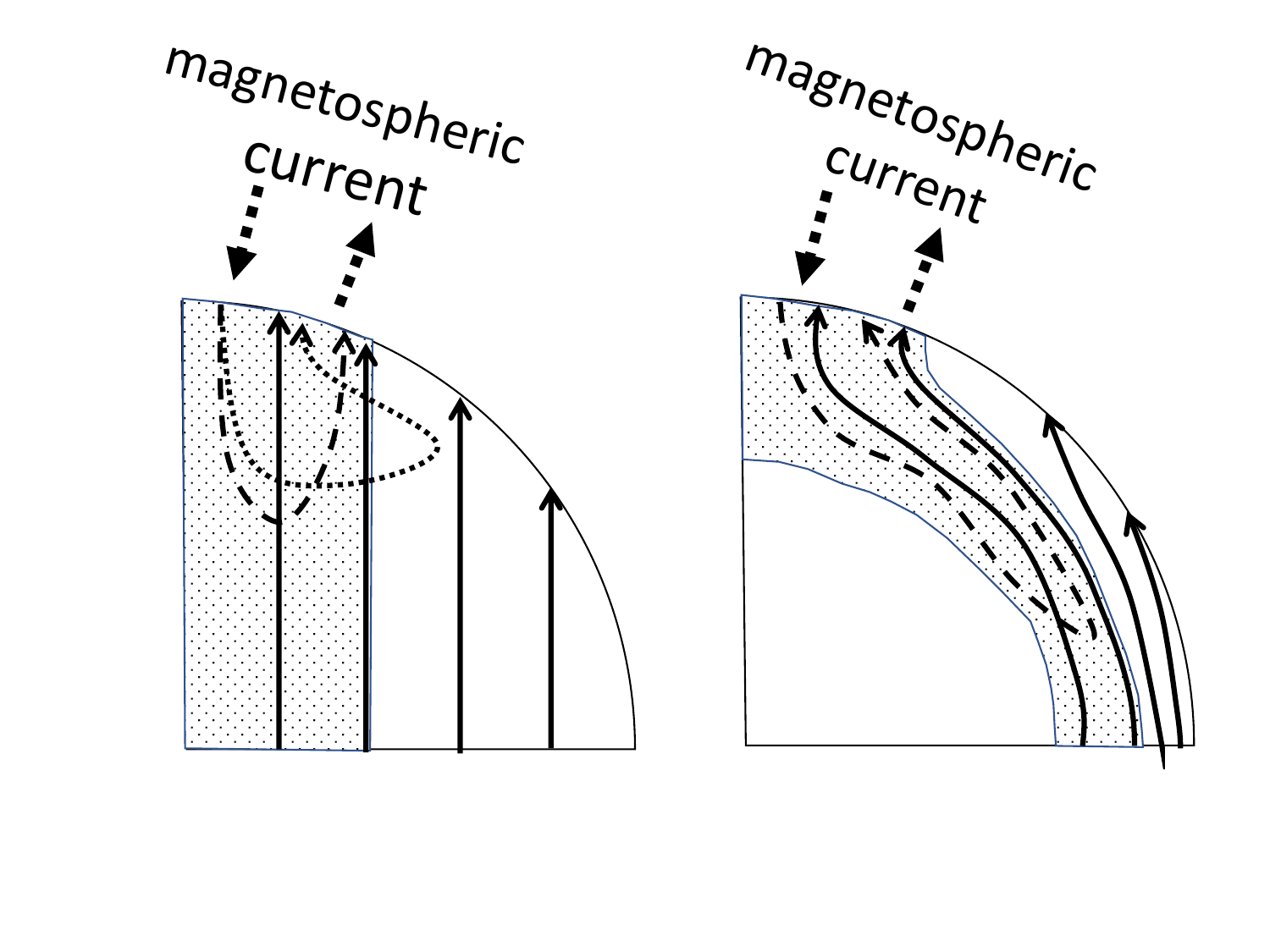}
\caption{ \label{torq.b-fig21} 
Example distributions of the magnetospheric current, depending on
the configuration of the polar cap magnetic flux inside the star.
The current lines  lie only within the polar cap magnetic flux (dashed curves).
The current lines penetrate outside of it (the dotted curve in the left panel)
is unlikely. If the polar cap magnetic flux (shaded region) extends not to
the core region but to the equatorial region in the crust,
the electromagnetic force is likely to extend to the equatorial region
in such a way that the braking force  is  balanced  the inertial force.
}
\end{center}
\end{figure}

If the magnetic flux is uniform as is assumed in our numerical example,
the current is distributed
as shown in Fig.~\ref{torq.b-fig10}~(b) 
and the dashed curve in the left panel of
Fig.~\ref{torq.b-fig21}.
The current does not penetrate outside the polar magnetic flux; 
if it did,
 the electromagnetic force would be in opposite 
direction to the braking force, and as a result, 
 the elastic energy would be increased.
It must be noted here that the inertial force dominates in the equatorial region.
Therefore,
if the polar cap magnetic flux is bent towards the equator such as 
shown in the right panel of Fig.~\ref{torq.b-fig21}, 
then the current penetrate to the equatorial region, 
and the braking stress is much more relaxed.
We find that  the distribution of the braking torque 
depends  thoroughly on the
distribution of the polar cap magnetic flux inside the star.
The Hall effect by the magnetospheric current, in other words,
the drift current due to the braking force, has little effect
to the background magnetic field inside the star.
Therefore, the key factor for the distribution of the braking force is 
the background magnetic field structure, for which the magneto-thermal evolution
has been intensively investigated 
(see e.g., \citet{2019LRCA....5....3P}).

The distribution of the braking
force will be calculated
if the magnetic field structure inside the star is given, and 
thereafter one can discuss the angular momentum
transfer inside the star.
The spin-down evolution including the glitches and any other kinds of timing
noise was discussed with the two-component, crust-superfluid, models
\citep{1969Natur.224..872B,2015IJMPD..2430008H,2021MNRAS.502.3113M}, where
the torque distribution inside the star was not taken into account. 
Our present study demonstrates that 
 the braking torque distribution 
for a given magnetic field can be calculated and hence
a more detailed modelling of the spin-down history than ever conducted is  possible.
Furthermore, the resultant models can
  be compared with
observations of  timing irregularities of pulsars to 
gain deeper insight about 
the magnetic structure inside the neutron star.

\section*{Acknowledgements}

This work is supported by KAKENHI 18H01246 (SS, SK),
19K14712 and 21H01078 (SK).
The authors thank the anonymous referee for comments which led to
improvements in the manuscript.

\section*{Data Availability}

The data underlying this article will be
shared on reasonable request to the corresponding author.



\appendix
\section{Appendix}
\subsection{Useful formulae for the obliquely rotating system}
In the steady rotating system with the angular velocity $\vec{\Omega}_*$,
the time-derivative can be eliminated;
for any scalar and vector fields, $\xi$ and $\vec{b}$, we have
\begin{equation}
\left[ {D \xi  \over Dt} \right]_{\rm rot}
\equiv \left( {\partial \over \partial t} 
  + \Omega_* {\partial \over \partial \varphi} \right) 
   \xi
=0,
\label{torq.c-st03}
\end{equation}
and
\begin{equation}
\left[ {D \vec{b} \over Dt} \right]_{\rm rot}
\equiv \left( {\partial \over \partial t} 
  + \Omega_* {\partial \over \partial \varphi} 
  - \vec{\Omega}_* \times \right)
   \vec{b}
=0,
\label{torq.c-st04}
\end{equation}
where $\Omega_* = |\vec{\Omega}_*|$, and 
$\varphi$ is the azimuthal angle.
The corotation vector defined by $\vec{u}_{\rm c} = \vec{\Omega}_* \times \vec{r}$
$= \Omega_* r \vec{e}_\varphi$
is useful, where $ r $ is the axial distance and 
$\vec{e}_\varphi$ is the unit toroidal vector.
For example,  
$\vec{u}_{\rm c} \cdot \nabla = \Omega_* \partial / \partial \varphi$,
$\nabla \cdot \vec{u}_{\rm c} =0$, 
$ \nabla \times \vec{u}_{\rm c} = 2 \vec{\Omega}_*$,
$(\vec{b} \cdot \nabla ) \vec{u}_{\rm c} = \vec{\Omega}_* \times {\bf b}$,
and with the help of the steadiness condition, we have
\begin{equation}
\nabla (\vec{u}_{\rm c} \cdot \vec{b} )
= - { \partial \vec{b} \over \partial t } 
  + \vec{u}_{\rm c} \times ( \nabla \times \vec{b}),
\label{torq.c-fml2}
\end{equation}
and 
\begin{equation} \label{torq.c-aid2}
\nabla \times (\vec{u}_{\rm c} \times \vec{b} )
= {\partial \vec{b} \over \partial t}
+ \vec{u}_{\rm c} (\nabla \cdot \vec{b} ).
\end{equation}

\bsp	
\label{lastpage}
\end{document}